\documentclass[aps,pra,superscriptaddress,showpacs,12pt]{revtex4}

\usepackage{amsmath}
\usepackage{amsfonts}
\usepackage{amssymb}
\usepackage{multirow}
\usepackage{verbatim}
\usepackage{alltt}
\usepackage{moreverb}
\usepackage{graphicx,color,graphics}
\usepackage{hyperref}
\usepackage{setspace}
\usepackage{url}

\newcommand{\cM}{{\cal M}}
\newcommand{\cO}{{\cal O}}
\newcommand{\cR}{{\cal R}}

\newcommand{\cT}{{\cal T}}
\newcommand{\LR}{\textrm{LR}}
\newcommand{\CH}{\textrm{CH}}
\newcommand{\logp}{\log\textrm{-}p}
\newcommand{\dom}{\textrm{dom}}
\newcommand{\rls}{\mathbb{R}}
\newcommand{\nats}{\mathbb{N}}
\newcommand{\proof}{\paragraph*{Proof.}}
\newtheorem{Theorem}{Theorem}
\newcommand{\qed}{\hspace*{\fill}\rule{2.5mm}{2.5mm}%
\vspace*{8pt}\par}

\providecommand{\ignore}[1]{}

\newcommand{\qvbar}{\mbox{$|\hspace*{-3pt}|\hspace*{-3pt}|$}}
\newcommand{\qrangle}{\mbox{$\rangle\hspace*{-4.3pt}\rangle\hspace*{-4.3pt}\rangle$}}

\newcommand{\ket}[1]{\qvbar{#1}\qrangle}

\newcommand{\lr}{\textrm{lr}}
\newcommand{\pr}{\textrm{pr}}

\begin{document}
\title{Bell Inequalities for Continuously Emitting Sources}
\author{Emanuel Knill}
  \email[Electronic address: ]{emanuel.knill@nist.gov}
\author{Scott Glancy}
\author{Sae Woo Nam}
\author{Kevin Coakley}
\affiliation{National Institute of Standards and Technology, Boulder, Colorado, 80305, USA}
\author{Yanbao Zhang}
\altaffiliation[Current address: ]{Institute for Quantum Computing, University of Waterloo, Waterloo, Ontario N2L 3G1, Canada}
\affiliation{Department of Physics, University of Colorado Boulder, Boulder, Colorado, 80309, USA}
\affiliation{National Institute of Standards and Technology, Boulder, Colorado, 80305, USA}

\begin{abstract}
  A common experimental strategy for demonstrating non-classical
  correlations is to show violation of a Bell inequality by measuring
  a continuously emitted stream of entangled photon pairs.  The
  measurements involve the detection of photons by two spatially
  separated parties. The detection times are recorded and compared to
  quantify the violation. The violation critically depends on
  determining which detections are coincident. Because the recorded
  detection times have ``jitter'', coincidences cannot be inferred
  perfectly.  In the presence of settings-dependent timing errors,
  this can allow a local-realistic system to show apparent
  violation--the so-called ``coincidence loophole''.

  Here we introduce a family of Bell inequalities based on signed,
  directed distances between the parties' sequences of recorded
  timetags. Given that the timetags are recorded for synchronized,
  fixed observation periods and that the settings choices are random
  and independent of the source, violation of these inequalities
  unambiguously shows non-classical correlations violating local
  realism.  Distance-based Bell inequalities are generally useful for
  two-party configurations where the effective size of the measurement
  outcome space is large or infinite. We show how to systematically
  modify the underlying Bell functions to improve the signal to noise
  ratio and to quantify the significance of the violation.
\end{abstract}


\maketitle

\section{Introduction}

Quantum mechanical systems can give rise to measurement correlations
that local realistic (LR) systems are unable to produce.  Physical
theories that satisfy the principle of local realism (LR) posit a set
of hidden variables associated with the physical systems.  The hidden
variables cannot be influenced by spacelike-separated events.  They
are not observable, but they determine the outcomes of all
measurements.  The existence of hidden variables constrains the
probability distributions that can describe LR systems.  In 1964 Bell
constructed an inequality that is satisfied by all correlations
accessible by LR and showed that correlations between
spacelike-separated measurements on two quantum systems can violate
this inequality~\cite{bell:qc1964a}.  The realization that quantum
mechanics allows more general probability distributions than LR has
motivated many experimental tests that have shown Bell-inequality
violations (see Ref.~\cite{genovese:qc2005a} for a review).  In
addition to the fundamental importance of tests of LR, systems
violating LR can be used for quantum information tasks such as quantum
key distribution~\cite{barrett:qc2005a, masanes:qc2009a,
  masanes:qc2011a} and secure randomness
generation~\cite{pironio:qc2010a, colbeck:qc2011a, vazirani:qc2012a}.
For these cryptographic applications, one must demonstrate violation
of LR with high statistical significance in the presence of
adversarial effects, such as a hacker who has tampered with the system
in an attempt to learn a secret key.

So far, all tests of LR have invoked additional assumptions about the
types of LR theories governing their experiments.  Examples include
the assumption that photon detection probabilities are not correlated
with the measurement choices or photon polarizations (the ``fair
sampling'' assumption), the assumption that measurement choices at one
location cannot influence events at another location even when they
are not spacelike separated, and the assumption that the sequence of
measurement choices and outcomes in an experiment are independent and
identically distributed (i.i.d.). For a review see
\cite{larsson:qc2014a}.  Various experiments have been able to relax
some of these assumptions, but no single experiment has been able to
reject the most general LR theories.  Due to recent advances in
entangled photon generation and photon detection, we anticipate that
an optical experiment that is free of additional assumptions will be
accomplished in the near future.

To test a Bell inequality in an experiment, one repeats the
preparation and joint measurement of spatially separated systems some
finite number $N$ times.  We call each such repetition a ``trial'' of
the experiment.  During a trial, entangled systems are sent to two or
more measurement locations.  At each location, a random choice is made
that determines which property will be measured.  Ideally, each trial
is clearly identifiable so that measurement choices and outcomes at
the various locations can be matched with one another.  However, in
many experiments this trial identification cannot be achieved with
perfect certainty.  A popular experiment design involves the
continuous pumping of a nonlinear crystal that produces photon pairs
through spontaneous parametric down-conversion.  In these experiments,
the entangled pairs are produced randomly in time.  Furthermore, the
detectors used at the two measurement locations (``Alice'' and
``Bob'') have nonzero timing jitter.  These effects can create
confusion about which events at Alice correspond to which events at
Bob.

To resolve this confusion one typically defines a ``coincidence
window'' in time, so that if the period of time between photon
detections at Alice and Bob is less than the coincidence window width,
the two events are considered to be part of the same trial.  If Alice
or Bob observe multiple detections within one coincidence window,
more sophisticated algorithms can be used to attempt to match Alice's
and Bob's detections. The choice of the coincidence window width
depends on balancing the expected inter-arrival time between entangled
pair creations and the detector timing jitter, so that the probability
for a trial to contain multiple photon pairs is small, and the
probability for a photon to be lost by falling outside of the
coincidence window is also small.

Unfortunately, local realistic theories or hackers can use this
uncertainty about trial identification to produce apparent violation
of Bell inequalities.  The photons being measured could have
correlations between their measurement outcomes and arrival times.
Correlations could also exist between multiple photon pairs detected
during the same coincidence window.  Previous tests of LR that use
continuously pumped spontaneous parametric down-conversion have (often
implicitly) assumed that such correlations do not exist.  Therefore
these experiments do not reject the most general LR theories; they
only reject LR theories that do not allow timing or cross-pair
correlations.  Experiments requiring these assumptions are said to
suffer from the ``coincidence loophole''.  Larsson and Gill described
local realistic theories that exploit the coincidence loophole and
methods to defeat them in~\cite{larsson:qc2004a}.  Their closure of
this loophole requires a bound on the probability that a true
coincidence is missed. This bound cannot be measured without
additional assumptions and must be trusted.

In this paper we describe a different method for closing the
coincidence loophole that does not require additional assumptions
about the systems being measured.  Instead, we define a ``trial'' as
all events occurring within a predetermined time interval.  The data
produced by the trial are the measurement choices at Alice and Bob and
the lists of times at which Alice and Bob detected photons (the
``timetags'').  The trial may be many times longer than the expected
time between photon pair creations, and Alice and Bob may observe many
photon detection events during one trial.  (In practice the length of
a trial will be constrained if one desires to achieve spacelike
separation between measurement choices on one side and detections on
the other.)

To test LR with such trial data, we develop distance-based Bell
inequalities.  Signed, directed distances between trial outcomes that
measure the dissimilarity between Alice's and Bob's lists of timetags
are defined such that they obey a directed triangle inequality.  These
distances are closely related to edit distances used to compare words
in spell checking or to align DNA sequences in computational biology.
Using LR to compute the expectation value of sums of these distances
yields an inequality satisfied by all LR theories.  These
distance-based Bell inequalities provide a rigorous analysis of tests
of LR based on continuously emitting sources; they enable the
rejection of a larger class of LR theories than previously known
methods.  Also, the triangle inequality can be a powerful tool for
finding new Bell inequalities in other contexts.  For example, see the
works of Dzhafarov and Kujala~\cite{dzhafarov:qc2013a} and of
Kurzynski and Kaslikowsi~\cite{kurzynski:qc2014a}.

In Sect.~\ref{sec:preliminaries} we introduce basic notions and define
relevant mathematical notation.  In Sect.~\ref{sec:tt_bell_tests} we
explain the experimental setup of ``timetag Bell tests'' in more
detail and describe a simple LR model that exploits the coincidence
loophole for an apparent violation of a Bell inequality.  In
Sect.~\ref{sec:distance} we define ``distance'' functions between
measurement outcomes that obey the directed triangle inequality.  We
then use the triangle inequality to derive Bell inequalities satisfied
by any LR theory regardless of the choice of distance function.  In
Sect.~\ref{sec:belltimetagseq} we describe functions for computing the
distance between timetag sequences and obtain the associated Bell
inequalities.  In Sect.~\ref{sec:nosigadj} we introduce non-signaling
equalities that constrain all theories that prohibit Alice from
sending information to Bob by use of her measurement choice (and
vice-versa).  Although both quantum and LR theories obey the
non-signaling equalities, these equalities can be used to transform
Bell inequalities and improve the signal-to-noise ratio (SNR) of the
inequalities' violation in an experiment.  In
Sect.~\ref{sec:analysis_protocol} we provide a protocol that sets
aside an initial segment of the data as a training set to determine a
good distance function.  In Sect.~\ref{sec:pbr} we discuss the
relationship between the SNR for the violation of a Bell inequality
and $p$-value bounds for rejecting LR.  Bounds on $p$-values can be
computed with Gill's martingale-based
protocol~\cite{gill:qc2001a,gill:qc2003a} or the prediction-based
ratio (PBR) protocol~\cite{zhang_y:qc2011a, zhang_y:qc2013a}.  The
main result of this section is a method for truncating distance
functions to enable application of these protocols.  The technique is
general and can be used on any Bell function derived from a triangle
inequality. Here, a useful step is to balance the violation between
the measurement settings by means of the non-signaling equalities.  In
Sect.~\ref{sec:sim} we apply timetag Bell inequalities to simulated
data.  We discuss the effects of detector inefficiency and detector
jitter on the violation of timetag Bell inequalities and on the
$p$-value bounds computed with the PBR protocol. We quantify the
violation and $p$-value bounds as functions of the jitter
distribution's width and quote lower bounds on the maximum jitter
width at which violation can be observed for photon detection
efficiencies ranging from $0.74$ to $0.95$. The simulations include an
LR model that exploits the coincidence loophole while closely
mimicking the measurement statistics of a Poisson source of entangled
photons measured with jittery detectors.  In the Appendix we describe
numerical methods to optimize parameters of the distance functions to
give high inequality violation, to compute distances for timetag Bell
inequalities, and to compute the SNR of the violation of an
inequality.  The Appendix also contains further details of the
coincidence-loophole-exploiting LR model.

\section{Preliminaries}
\label{sec:preliminaries}

We consider experiments to test LR, where an experiment consists of a
sequence of trials. The trials' measurement outcomes need not be
independent from one trial to the next, but before the next trial,
there is a probabilistic description of the next trial's outcome,
where the probabilities may depend on the past and current
conditions. The class of LR models of interest is defined by
specifying constraints on these probabilities.  We consider the case
where a trial consists of observations by two parties, $A$ and $B$,
each of whom can choose one of two measurement settings for their
observation. We leave extensions to more parties and settings for
future work. The full trial outcome includes the settings chosen as well as
the measurement outcomes. In many cases, the measurement outcomes are
two-valued. For example, the outcome may indicate whether a
photodetector ``clicked'' or not.  Here we consider arbitrary outcome
spaces, but focus on the case where a party's measurement outcome
is an ordered sequence of timetags of events, for example detection
events. Thus, there is no bound on the size of the outcome space

We denote the random variable for a trial's outcome including the
settings by $T$. This random variable is a tuple of four random
variables $T=(O^A,S^A,O^B,S^B)$ where $O^X$ is $X$'s measurement
outcome and $S^X$ is $X$'s chosen setting. The two possible settings
are denoted by $\bar 1$ and $\bar 2$.  We also use the notation
$T^X=(O^X,S^X)$. We follow the notational convention that random
variables ($O,S,T,\ldots$) are denoted by roman upper case
letters. This is also true of party labels, but the distinction should
be clear from context.  The range of random variable $R$ is denoted by
$\cR$. Observed values of random variables are denoted by their
corresponding lower-case letters; for example, $r$ denotes an observed
value of $R$.  Superscripts and subscripts serve to identify members
of a family of conceptually related random variables or to select out
parts of tuple-valued random variables.  Formally, the random
variables for a trial are functions on an underlying probability space
that includes any ``hidden'' variables that may play a role, but we do
not need to explicitly refer to this space here.

A deterministic LR model must, before a trial and independent of the
settings, commit to a specific measurement outcome $d^{X}_c$ for each
party $X=A,B$ and each setting $c=\bar 1,\bar 2$.  A general LR model
is a probabilistic mixture of deterministic models.  (One can imagine
that a hidden random variable selects which deterministic model
controls a trial's measurement outcomes.) Thus, an LR model is
described by a random variable $D_{\LR} = (D^A_{\bar 1},D^A_{\bar
  2},D^B_{\bar1},D^B_{\bar 2})$, where $T$ relates to $D_{\LR}$
according to $T=(D^A_{S^A},S^A,D^B_{S^B},S^B)$.  Although the
parties cannot simultaneously measure both settings $\bar1$ and
$\bar2$ in a single trial, the LR model allows for that possibility by
pre-assigning measurement outcomes to both settings.  That such a
pre-assignment exists is the essential claim of realism.  Quantum
theory does not pre-assign outcomes and disallows the possibility that
the two settings can be measured simultaneously. Quantum theory can
thereby achieve a larger set of trial probability distributions.

In an idealized test of LR, the settings choices are made randomly and
independently of $D_{\LR}$ according to a probability distribution
that is under experimenter control. In this case, LR models satisfy
that $S=(S^A,S^B)$ is independent of $D$, and the probability
distribution of $S$ is known before the trial. This defines LR models
satisfying the free choice assumption. For the remainder of the paper,
LR models are assumed to satisfy free choice, and by default the
settings distribution is uniform.

From a mathematical and statistical point of view, a successful test
of LR shows that probabilistic LR models are statistically
inconsistent with the data. Interpretation of the inconsistency
requires additional analysis and can depend on the experimental
context. In fact such interpretations could attribute the
inconsistency to the presence of so-called ``loopholes'' rather than
to the falsity of LR. An experimental goal is to convincingly exclude
the presence of such loopholes.

\section{Timetag Bell tests}
\label{sec:tt_bell_tests}

A common method for performing Bell tests is to use a source that
continuously emits pairs of polarization entangled photons.  The
photons are delivered to two measurement setups. A trial consists of
choosing the settings and then recording photodetection events for a
fixed observation window. We focus on the simplest case, where the
measurement setups involve polarizers whose angles determine the
settings. Each setup has one photodetector that records photons that
passed through the polarizer.  Thus, the record of a trial includes
two timetag sequences recording the times at which photons were
detected. The experiments reported in~\cite{giustina:qc2013a,
  christensen:qc2013a} used this setup.

One way to think about such an experiment is that fundamentally, each
photon pair's emission and detection constitutes a trial.  In this
case, the first step in an analysis is to identify the detection
pattern for each emitted pair. The record does not identify when
neither photon was detected, but the Bell inequalities used can be
chosen so that the total Bell-inequality violation is insensitive to
the number of photon-pair trials where neither photon was detected.
Thus, the analysis requires identifying coincidences, that is, pairs
of detections that are due to one photon pair. Identifying
coincidences is complicated by the fact that the recorded timetags
have ``jitter'', that is, the difference between the timetag $t$ and
the ``true'' time of arrival of a photon $t_0$ is a random variable
with non-negligible width $j$ (to be defined in Sect.~\ref{sec:sim}
for specific jitter distributions). Furthermore, since photon pairs
are continuously emitted, their creation times and their times of
arrival are also random. Pair emission can usually be modeled as a
Poisson process.  Denote the mean inter-arrival time between
successive photon pairs as $\tau$.  It is necessary to determine which
pairs of close timetags $t$ and $r$ of $A$ and $B$ are due to the same
photon pair.  This cannot be done without error, as there is always
the possibility that photons detected by $A$ and $B$ around the
same time are from two different photon pairs that were created with
small time separation. The probability of this event grows with
$j/\tau$.

Given that coincidences cannot be identified exactly, it is necessary
to determine how this affects the interpretation of a Bell-inequality
violation. In cases where the nominal mean violation per photon pair
is small and $j/\tau$ is relatively large, the evidence against LR may
be weakened substantially. An example of this situation is the
experiment reported in Ref.~\cite{giustina:qc2013a}, which aimed to
close the fair sampling loophole with photons.  The violation was
limited by the overall detection efficiency realized in the
experiment.  To interpret the violation, one can analyze the effects
of coincidence identification error by making the assumptions that the
source is idealized Poisson, the jitter is settings independent, the
photon pairs' states are identical and independent, and the method for
recording detections is memoryless. Parts of such an analysis are
in~\cite{kofler:qc2013a}. But these are highly idealizing assumptions
unlikely to be satisfied in a real experiment. Relying on them
precludes making strong claims on having demonstrated non-LR
effects. Of particular concern is that the presence of jitter in
combination with a conventional coincidence analysis requires a fair
coincidence sampling assumption~\cite{larsson:qc2004a}.  For
conventional analyses, this assumption can be avoided by using
``pulsed'' trials, which can significantly reduce the rate of
detections. Such an experiment was reported in
Ref.~\cite{christensen:qc2013a}.

\begin{figure}
\includegraphics[width=3in]{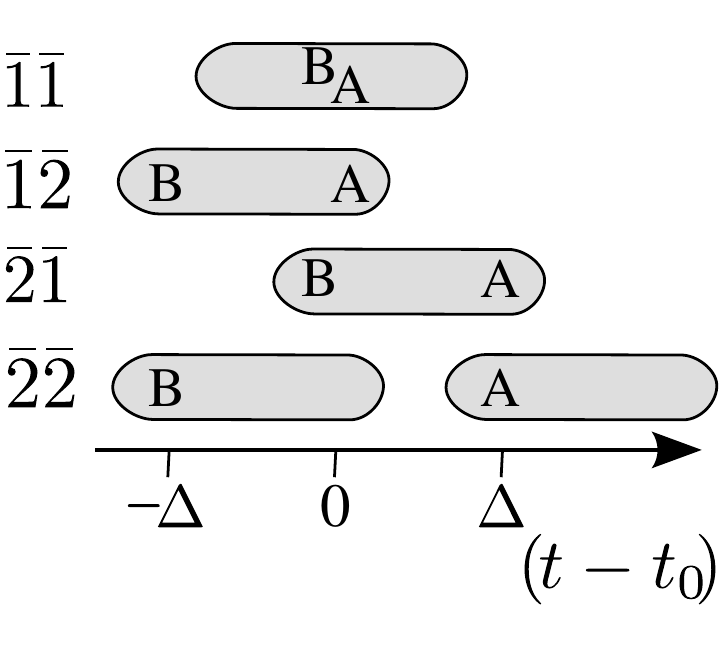}
\caption{\label{fig:coincidence_loophole} Illustration of an LR model
  that uses setting dependent detection times to violate a Bell
  inequality. Detection time $t-t_0$ proceeds along the horizontal
  axis.  Each row above the time axis corresponds to one measurement
  setting combination used by $A$ and $B$.  Photons are detected by
  $A$ and $B$ at the times labeled ``A'' and ``B''.  (At the
  $\bar1\bar1$ setting, both $A$ and $B$ detect their respective
  photons near $0$.)  The shaded regions indicate coincidence
  windows of width $1.5\Delta$.  At the $\bar2\bar2$ setting, the
  photons detection times are separated by $2\Delta$, so $A$ and $B$ can
  never observe coincidences with $\bar2\bar2$.}
\end{figure}

Unfair coincidence sampling can arise from local, settings-dependent
properties of the detectors, including the associated settings-related
apparatus such as polarization filters.  A simple LR model that
exploits unfair coincidence sampling to show violation of a Bell
inequality is illustrated in Fig.~\ref{fig:coincidence_loophole}.
Suppose that for $A$, the difference $t-t_0$ between the recorded
timetag and the photon arrival time is $0$ on setting $\bar1$ and
$\Delta$ on setting $\bar2$. For $B$, suppose that the difference is
$0$ on setting $\bar1$ and $-\Delta$ on setting $\bar2$. To identify
coincidences, one can choose a coincidence window width $w$ and
declare that timetags of $A$ and $B$ whose separation is less than $w$
are coincident. It is necessary to have a method for resolving
coincidence conflicts such as when a timetag of $A$ is within $w$ of
more than one timetag of $B$. Here we just assume that $j/\tau$ is
small enough for this not to be considered an issue.  Suppose that the
coincidence window width is chosen to be $w=1.5\Delta$.  If the LR model for
the photon pair is to ``detect'' no matter what the setting is, then
$A$ and $B$ record coincidences on all settings except $\bar2\bar2$,
where the two detections are inferred as being non-coincidence
detections. This LR model strongly violates commonly used Bell
inequalities, such as the one introduced in Sect.~\ref{sec:distance},
Eq.~\eqref{eq:CHSH2}, whose violation is increased by anticorrelation
between $A$'s and $B$'s detections on the $\bar2\bar2$ setting.  The
choice of $w$ may seem arbitrary, but a natural way to choose $w$ is
to optimize the violation on a training set or on preliminary data.
In this case, $w=1.5\Delta$ is an optimal choice.  Note that locality
assumptions and the assumption that settings-choices are uniformly
random and independent of hidden variables affecting the measurement
outcomes are satisfied in this example.

A detailed theoretical treatment of unfair coincidence sampling is
given in Ref.~\cite{larsson:qc2004a}, including a more sophisticated
example that can respond to the continuous angular settings choices
available when measuring photons.  Given the assumptions in
Ref.~\cite{larsson:qc2004a}, there are valid adjustments to a Bell
inequality based on knowledge of the probability of missing a
coincidence. Our approach based on timetag Bell inequalities does not
require such additional knowledge.

It may seem like the presence of unfair coincidence sampling due to a
dependence of detector timing on settings can be excluded by checking
that there is no widening of the time-separation distribution of the
nearest timetags of $A$ and $B$ for the $\bar2\bar2$ settings
compared to the others.  Further, one can attempt to choose $w$ after
studying this distribution to ensure that the fraction of missed
coincidences for the $\bar2\bar2$ setting is sufficiently small.  Any such
attempt would require an hypothesis test (or some other way to
quantify evidence) for the claim that no unfair coincidence sampling
is present in the experiment.  Depending on how the data for these
tests is acquired, additional assumptions on consistency of detector
behavior may be needed.  In pursuing this approach, one must then
decide at what significance level one wishes to exclude excessive
unfair coincidence sampling.  This significance level should be
similar to the claimed significance of the violation.  This may not be
feasible in an experiment without greatly weakening the significance
of the result.

Coincidence sampling effects can also be exploited directly by an LR
source. The simplest example just simulates the detector timing issue.
There are two operationally different versions of this example. In the
first, the local hidden variable of a photon also identifies the time
at which the detector records it in a setting-dependent way. One could
imagine that once the photon arrives at the detector and senses the
setting, it ``pauses'' a variable amount of time.  It now suffices to
emit photon pairs where photon $A$'s local variable assigns ``detect''
at time $t$ on setting $\bar1$ and ``detect'' at time $t+\Delta$ on
setting $\bar2$, whereas photon $B$'s variable assigns ``detect'' at
time $t$ on setting $\bar1$ and ``detect'' at time $t-\Delta$ on
setting $\bar2$.  Such photons seem impossible to realize, but the
following version of the example may be realizable. The source sends a
photon to $B$ at time $t-\Delta$, a pair to $A$ and $B$ at time $t$,
and another to $A$ at time $t+\Delta$.  The first photon is prepared
so as to be detected by $B$ only on setting $\bar2$.  The two middle
photons are detected only on setting $\bar1$.  The last photon is
prepared so as to be detected by $A$ only on setting $\bar2$. When $A$
and $B$ compare their timetags and use $w=1.5\Delta$ for their
coincidence analysis, they again see coincidences on all settings
except for $\bar2\bar2$. An LR source that wishes to hide having
manipulated emission times can intersperse a small number of photons
with the emission pattern above with regular LR pairs of photons that
ensure equality for the Bell inequality of interest. Other
opportunities to hide the presence of unfair coincidence sampling from
the experimenters exist. The LR source can systematically introduce LR
photons with varying timing features, and it can conditionally omit
``normal'' photon pairs to make it more difficult to see excess
numbers of close detections outside the coincidence window. While
these possibilities may seem physically unrealistic, they are of
concern in cryptographic applications of experimental configurations
for violating Bell inequalities. To show that these concerns are
justified, in Sect.~\ref{sec:sim} we show simulation results for an LR
source whose statistics would be difficult to distinguish from a
quantum source in an experimental setting. The conventional
coincidence analysis shows a false violation of a Bell inequality for
this source.

\section{Distance-based Bell functions}
\label{sec:distance}

\begin{figure}
\includegraphics[width=3in]{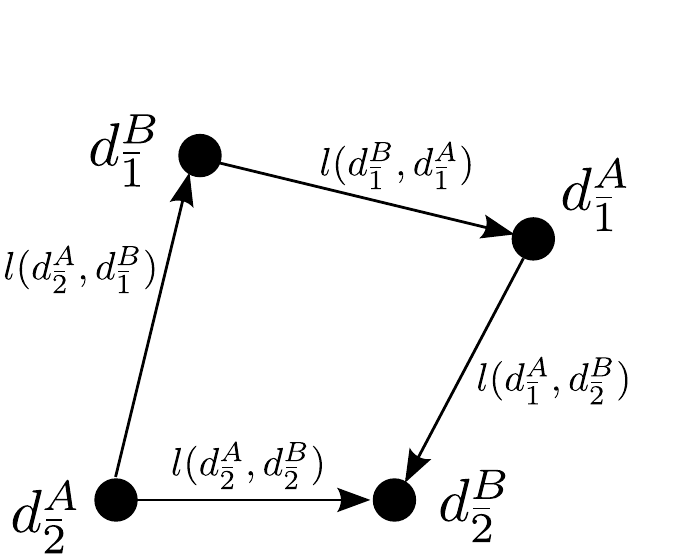}
\caption{\label{fig:iterated_triangle}Illustration of the
  twice-iterated triangle inequality.  Each node represents one of the
  four potential measurement outcomes in $d_{\LR}$, and each directed
  edge represents one of the lengths $l$ in
  Eq.~\ref{eq:iterated_triangle}.  The edges on the path correspond to
  ``compatible'' measurements pairs that are experimentally
  measured. The locations of the nodes are arbitrary in this
  illustration.}
\end{figure}

Distance-based Bell functions and associated Bell inequalities
generalize the conventional two-party, two-setting Bell inequalities
such as the CHSH~\cite{clauser:qc1969a} and CH
inequalities~\cite{clauser:qc1974a} (the abbreviations
stand for the authors' initials).  Consider two parties $A$ and $B$,
where each can choose from two settings labeled $\bar 1$ and $\bar 2$
and the physical meaning of the setting labels depends on the
party. Suppose that the measurement outcomes at a given setting are
$-1$ or $1$. Let $O^X$ and $S^X$ be the measurement outcome and
setting of party $X$, respectively.  One of the CHSH inequalities for this
configuration is
\begin{eqnarray}
\langle O^AO^B|S^A=\bar2,S^B=\bar2\rangle\hspace*{.25in}&&\nonumber\\
{}-\langle O^AO^B|S^A=\bar2,S^B=\bar1\rangle&&\nonumber\\
{}-\langle O^AO^B|S^A=\bar1,S^B=\bar1\rangle&&\nonumber\\
{}-\langle O^AO^B|S^A=\bar1,S^B=\bar2\rangle
  &\geq& -2,
\label{eq:CHSH1}
\end{eqnarray} 
where $\langle U\rangle$ denotes the expectation of $U$.  This
inequality is satisfied by all LR models. Let
$l(x,y) = |x-y|$.
Since $O^AO^B = 1-l(O^A,O^B)$, Eq.~\eqref{eq:CHSH1} can be rewritten
as
\begin{eqnarray}
\langle l(O^A,O^B)|S^A=\bar2,S^B=\bar1\rangle\hspace*{.25in}&&\nonumber\\
{}+\langle l(O^B,O^A)|S^A=\bar1,S^B=\bar1\rangle&&\nonumber\\
{}+\langle l(O^A,O^B)|S^A=\bar1,S^B=\bar2\rangle&&\nonumber\\
{}-\langle l(O^A,O^B)|S^A=\bar2,S^B=\bar2\rangle
  &\geq& 0.
\label{eq:CHSH2}
\end{eqnarray} 
A deterministic LR model assigns a specific value $d^X_c$ for the
measurement outcome of each party $X$ and setting $c$ before the
experiment. In this case, the left-hand side of Eq.~\eqref{eq:CHSH2}
is given by
$l(d^A_{\bar2},d^B_{\bar1})+l(d^B_{\bar1},d^A_{\bar1})+l(d^A_{\bar1},d^B_{\bar2})-l(d^A_{\bar2},d^B_{\bar2})$.
This is at least $0$ because $l$ satisfies the triangle inequality as
illustrated in Fig.~\ref{fig:iterated_triangle}. The inequality of
Eq.~\eqref{eq:CHSH2} follows because general LR models are
probabilistic mixtures of deterministic ones.  We have switched the
$A$ and $B$ arguments of $l$ in the contribution for the $\bar1\bar1$
setting in preparation for applying asymmetric functions $l$.

The Bell inequalities of Eq.~\eqref{eq:CHSH1} and~\eqref{eq:CHSH2} are
expressed in terms of quantities that are conditional on 
settings. Since measurement settings need to be chosen randomly,
experimentally estimating these quantities requires dividing by the
actual number of times the relevant settings are chosen.  This
complicates the estimation of experimental uncertainty. To avoid this
complication, recall that the probability distribution of the settings
is known beforehand and is independent of the LR model. Let $p_{ab}$
be the probability that $A$ and $B$ choose settings $a$ and
$b$, respectively. Then the left-hand side of Eq.~\eqref{eq:CHSH2} is
equal to
\begin{equation}
\left\langle  (-1)^{[S^A=\bar2 \& S^{B}=\bar2]}l(O^A,O^B)/p_{S^AS^B}\right\rangle,
\label{eq:function_CHSH2}
\end{equation} 
where the expression $[\phi]$ in the exponent of $-1$ evaluates to $1$
when the logical formula $\phi$ is true and to $0$ when it is false.
Eq.~\eqref{eq:function_CHSH2} is the expectation of a function of the
settings and measurement outcomes. This function is the Bell function
for the inequality of Eq.~\eqref{eq:CHSH2}. For our default assumption
of a uniform settings probability distribution, $p_{S^{A}S^{B}}=1/4$.

In general, we define a Bell function to be a function of trial
outcomes whose expectation with respect to every LR model is
non-negative, where the probability distribution of the settings is
fixed.  We now obtain such Bell functions from functions
$l:(x,y)\in\cO\times\cO\mapsto l(x,y)\in \rls$ that satisfy the
twice-iterated triangle inequality $l(x_1,x_4)\leq
l(x_1,x_2)+l(x_2,x_3)+l(x_3,x_4)$, where $\cO$ is a common measurement
outcome space for all parties and settings.  Unlike a true distance,
we do not require $l$ to be non-negative or symmetric.  Consider a
deterministic LR model given as above by its outcome assignments
$d_{\LR}=(d^A_{\bar 1},d^A_{\bar 2},d^B_{\bar 1},d^B_{\bar 2})$.  The
elements of $d_{\LR}$ obey the inequality
\begin{equation}
0 \leq 
   l(d^A_{\bar2},d^B_{\bar1})+l(d^B_{\bar1},d^A_{\bar1})+l(d^A_{\bar1},d^B_{\bar2})
   - l(d^A_{\bar2},d^B_{\bar2}),
\label{eq:iterated_triangle}
\end{equation}
obtained by arranging the twice-iterated triangle inequality as
illustrated in Fig.~\ref{fig:iterated_triangle}.  The reversal of the
parties in the term $l(d^B_{\bar1},d^A_{\bar1})$ (the middle edge of
the indirect path) requires argument reversals in the expressions
below.

We can define a Bell function based on $l$ as follows:
\begin{equation}
B_l(t)=\left\{
  \begin{array}{ll}
     4l(o^B,o^A) & \mbox{if $s^A=\bar 1$ and $s^B=\bar 1$,}\\
     4l(o^A,o^B) & \mbox{if $s^A\not=s^B$,}\\
     -4l(o^A,o^B) & \mbox{if $s^A=\bar 2$ and $s^B=\bar 2$,}
  \end{array}\right.
\label{eq:chbelldef1}
\end{equation}
where $t=(o^A,s^A,o^B,s^B)$ is a trial outcome.  The factor of $4$
originates as the inverse of the probability $1/4$ of each of the
settings.  Consider an LR model with measurement outcome random
variables $D^X_c$. In the trial outcome random variable $T$,
$O^A=D^A_{S^A}$ and $O^B=D^B_{S^B}$. Because the settings are
independent of the $D^X_c$, the expectation of $B_l$ can be computed
as
\begin{equation}
  \langle B_l(T)\rangle_{\LR} =  
   \langle l(D^A_{\bar2},D^B_{\bar1})\rangle_{\LR} 
   +\langle l(D^B_{\bar1},D^A_{\bar1})\rangle_{\LR} 
   +\langle l(D^A_{\bar1},D^B_{\bar2})\rangle_{\LR} 
   - \langle l(D^A_{\bar2},D^B_{\bar2})\rangle_{\LR}  \geq 0,
\label{eq:bell_ineq}
\end{equation}
where the factors of $4$ were canceled by the probabilities of the
settings.  The inequality can be checked directly for
deterministic LR models, by replacing the random variable $D^X_c$ with
the constant $d^X_c$. For general LR models both sides can be
integrated with respect to the appropriate distributions over
deterministic LR models.  For a general settings
distribution, the values of $l$ in Eq.~\eqref{eq:chbelldef1} are
multiplied by the inverse of the applicable settings'
probability so that the expression for the expectation in
Eq.~\eqref{eq:bell_ineq} is unchanged.

Eq.~\eqref{eq:bell_ineq} is the Bell inequality associated with the
Bell function $B_l$. To test it in an experiment involving a sequence
$T_k$ of trial outcomes, one computes the values $B_l(t_k)$ on the
actual trial outcomes $t_k$. The sum $v=\sum_k B_l(t_k)$ is then an
estimate of $\sum_k \langle B_l(T_k)\rangle$, that can be compared to
$0$.  The test is considered successful if $v<0$, and the difference
betwen $v$ and $0$ is statistically significant.  According to the
conventional approach, this involves determining an uncertainty for
$v$.  See Appendix.~\ref{sect:nomsnr} for an effective method for
obtaining such an uncertainty that takes into account the known
probability distribution of the settings.  While this prescription is
seemingly straightforward, care must be taken when interpreting the
results for trials that may not be identical and
independent~\cite{gill:qc2001a,barrett_j:qc2002b,gill:qc2002a}.
Statements on the strength of evidence against LR require additional
analysis based on statistical hypothesis testing, see
Sect.~\ref{sec:pbr}.

For our applications to timetag sequences, we further generalize the
Bell functions by allowing $l$ to depend on the
settings. Consider functions $l_{ab}$ of two measurement outcomes that
satisfy the following version of the iterated triangle inequality:
\begin{equation}
  0 \leq 
  l_{\bar2\bar1}(d^A_{\bar2},d^B_{\bar1})+l_{\bar1\bar1}(d^B_{\bar1},d^A_{\bar1})+l_{\bar1\bar2}(d^A_{\bar1},d^B_{\bar2})
  - l_{\bar2\bar2}(d^A_{\bar2},d^B_{\bar2}).
\label{eq:iterated_triangle2}
\end{equation}
We call such an $l$ a $\CH$ function.  For the first and shorter
expression in the next definition, we use the notation $\tilde
l_{\bar1\bar1}(o_{1},o_{2}) = l_{\bar1\bar1}(o_{2},o_{1})$ and $\tilde
l_{ab}(o_{1},o_{2}) = l_{ab}(o_{1},o_{2})$ for $ab\not=\bar1\bar1$.  A
Bell function $B_l$ can now be defined by generalizing
Eq.~\eqref{eq:chbelldef1} according to
\begin{eqnarray}
B_l(t)&=& 
 4(-1)^{[s^{A}=\bar2 \& s^{B}=\bar2]} \tilde l_{s^{A}s^{B}}(o^{A},o^{B})\nonumber\\\nonumber\\
 &=&\left\{
  \begin{array}{ll}
     4l_{s^As^B}(o^B,o^A) & \mbox{if $s^A=\bar 1$ and $s^B=\bar 1$,}\\
     4l_{s^As^B}(o^A,o^B) & \mbox{if $s^A\not=s^B$,}\\
     -4l_{s^As^B}(o^A,o^B) & \mbox{if $s^A=\bar 2$ and $s^B=\bar 2$.}
  \end{array}\right.
\label{eq:chbelldef2}
\end{eqnarray}
We call $B_l$ a $\CH$ Bell function.  It satisfies
Eq.~\eqref{eq:bell_ineq} after adding the appropriate indices to the
occurrences of $l$.

We remark that every Bell function for a two-party, two-settings
configuration can be put in the form of a $\CH$ Bell function for
some choice of $\CH$ function $l$. Thus many of the techniques
discussed in the remainder of the paper are generally
applicable. Consider an arbitrary Bell function
$B(o^{A},s^{A},o^{B},s^{B})$ standardized as above so that $\langle
B(T)\rangle_{\LR}\geq 0$ for the settings probability distribution
$p_{ab}$. We can define
\begin{equation}
l_{B,ab}(o_{1},o_{2}) = \left\{
  \begin{array}{ll}
    B(o_{2},a,o_{1},b)p_{ab} & \mbox{if $ab=\bar1\bar1$,}\\
    B(o_{1},a,o_{2},b)p_{ab} & \mbox{if $a\not=b$,}\\
    -B(o_{1},a,o_{2},b)p_{ab} & \mbox{if $ab=\bar2\bar2$.}
  \end{array}
  \right.
\end{equation}
Since there are no constraints on the measurement outcomes in
deterministic $\LR$ models, the Bell inequality implies
the iterated triangle inequality of Eq.~\eqref{eq:iterated_triangle2}.
To show this, fix the LR model so that it is deterministic according
to $d_{\LR}$, where $d_{\LR}$ is arbitrary. Then
\begin{eqnarray*}
0&\leq& \langle B(T)\rangle_{\LR}\\
  &=& \sum_{ab}p_{ab}\langle B(O^{A},a,O^{B},b)|ab\rangle_{\LR}\\
  &=& \sum_{ab}p_{ab} B(d^{A}_{a},a, d^{B}_{b},b)\\
  &=& l_{B,\bar2\bar1}(d^{A}_{\bar2},d^{B}_{\bar1}) 
      + l_{B,\bar1\bar1}(d^{B}_{\bar1},d^{A}_{\bar1}) 
      + l_{B,\bar1\bar2}(d^{A}_{\bar1},d^{B}_{\bar2}) 
      - l_{B,\bar2\bar2}(d^{A}_{\bar2},d^{B}_{\bar2}).
\end{eqnarray*}

\section{Bell functions for timetag sequences}
\label{sec:belltimetagseq}

For our construction of $\CH$ functions for general timetag sequences,
we require functions $l_{ab}$ whose domains are two lists of real
numbers representing the timetag sequences at settings $ab$, whose
ranges are the real numbers, and that satisfy the iterated triangle
inequality.  Our general strategy is to construct $l_{ab}$ so that it
computes a quantity similar to an edit distance, which quantifies the
degree of dissimilarity between the timetag sequences obtained by
$A$ and $B$.  (However our $\CH$ functions may be negative and are not
symmetric in their arguments, so they are not strictly distances.)
Our implementation matches timetags of $A$ with timetags of $B$ and
assigns a cost to the difference between the matched timetags and a
cost to unmatched timetags.

To compute the cost for matched timetags we use function-tuples
$(f_{ab})_{a,b\in\{\bar1,\bar2\}}$, where the $f_{ab}:x\in\rls\mapsto
f_{ab}(x)\in\rls$ satisfy that for all $x,y,z\in\rls$,
$f_{\bar2\bar2}(x+y+z) \leq
f_{\bar2\bar1}(x)+f_{\bar1\bar1}(y)+f_{\bar1\bar2}(z)$. We denote the
set of such function-tuples by $\cT_4$.  The goal is to compare
timetags $r$ and $t$ obtained at settings $ab$ by computing
$f_{ab}(t-r)$.  To construct a function-tuple in $\cT_{4}$ from any
given $f_{\bar2\bar1}$, $f_{\bar1\bar1}$ and $f_{\bar1\bar2}$, we can
choose $f_{\bar2\bar2}$ such that $f_{\bar2\bar2}(u) \leq
\inf_{x,y\in\rls}
(f_{\bar2\bar1}(x)+f_{\bar1\bar1}(y)+f_{\bar1\bar2}(u-x-y))$, provided
the expression on the right-hand side is bounded from below. This
condition is satisfied if the given $f_{ab}$ are lower bounded.  Three
immediate examples of members of $\cT_4$ are the linear tuples with
$f_{ab}(x)=\lambda x$, the constant tuples with $f_{ab}(x) = c_{ab}$
where $c_{\bar2\bar1}+c_{\bar1\bar1}+c_{\bar1\bar2}= c_{\bar2\bar2}$
(an \emph{exact} and \emph{constant} function-tuple), and $f_{ab}(x) =
[x\geq 0]$. Here, we again used the convention that for a logical
formula $\phi$, $[\phi]=1$ if $\phi$ is true and $0$ otherwise.  To
construct other members of $\cT_4$ it helps to apply closure
properties of $\cT_4$.

\begin{Theorem}
The set of function-tuples $\cT_4$ is closed under the following operations.
\begin{itemize}
\item[(A)] Component-wise addition and multiplication by a positive real number
($\cT_4$ is a convex cone).
\item[(B)] The reflection defined by $f'_{ab}(x)=f_{ab}(-x)$.
\item[(C)] Component-wise maximum of two function-tuples.
\item[(D)] For real numbers $t_{ab}$ satisfying
  $t_{\bar2\bar2}=\sum_{ab\not=\bar2\bar2} t_{ab}$, the shift transforming the
  components according to $f_{ab}'(x)= f_{ab}(x+t_{ab})$.
\item[(E)] For non-negative function-tuples and $c\geq 0$, the 
  transformation defined by $f'_{ab}(x)=\min(f_{ab}(x),c)$.
\item[(F)] Let $f$ and $f'$ be function-tuples with $f'_{\bar2\bar2}$
  monotone non-decreasing. Then the function-tuple defined by
  $f''_{ab}=f'_{ab}\circ f_{ab}$ is a function-tuple. If
  $f_{\bar2\bar2}$ is also monotone non-decreasing then so is
  $f''_{\bar2\bar2}$.
\end{itemize}
\label{thm:ftuple_closure}
\end{Theorem}

\proof (A) can be checked by direct application of the definitions.
(C) follows from the observation that the maximum is monotone in each
argument and the maximum of two matched sums is at most the sum of the
maximums of the terms.  (B) and (D) follow from invariance of the
defining inequalities under reflection and under the shift specified
in (D).  To check (E) note that $f'_{\bar2\bar2}(x+y+z) \leq
f_{\bar2\bar2}(x+y+z)\leq
f_{\bar2\bar1}(x)+f_{\bar1\bar1}(y)+f_{\bar1\bar2}(z)$.  If a term on
the right-hand side is greater than $c$, non-negativity implies that
the right-hand side is at least $c$, an upper bound on the left-hand
side by definition of $f'_{\bar2\bar2}$. If not, the right-hand side
is equal to
$f'_{\bar2\bar1}(x)+f'_{\bar1\bar1}(y)+f'_{\bar1\bar2}(z)$.  
The following inequalities show (F):
\begin{eqnarray}
  f''_{\bar2\bar2}(x+y+z) &=&
  f'_{\bar2\bar2}(f_{\bar2\bar2}(x+y+z)) \nonumber\\
  &\leq& f'_{\bar2\bar2}(f_{\bar2\bar1}(x)+f_{\bar1\bar1}(y)+f_{\bar1\bar2}(z)) \nonumber\\
  &\leq& f'_{\bar2\bar1}(f_{\bar2\bar1}(x))+f'_{\bar1\bar1}(f_{\bar1\bar1}(y))+f'_{\bar1\bar2}(f_{\bar1\bar2}(z))\nonumber\\
&=& f''_{\bar2\bar1}(x)+f''_{\bar1\bar1}(y)+f''_{\bar1\bar2}(z),
\end{eqnarray}
and the composition of monotone non-decreasing functions is monotone
non-decreasing.
\qed

\begin{figure}
\includegraphics[width=\textwidth]{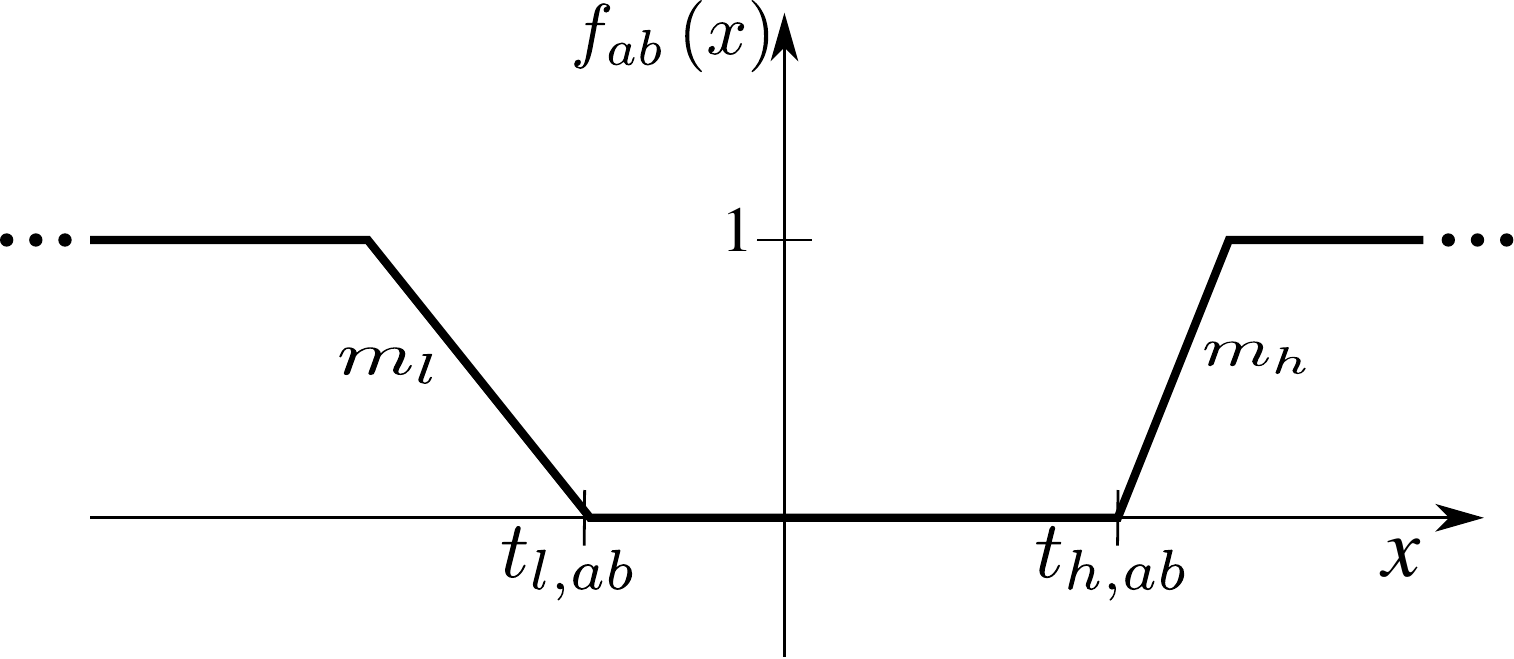}
\caption{\label{fig:linear-edge_window}Illustration of the linear-edge
  window functions. See the text for the definitions.  }
\end{figure}

Here are a few more examples of function-tuples in $\cT_4$.
\begin{itemize}
\item[(1)] $f_{ab}(x)=|x|=\max(x,-x)$. In this case, the condition is just the
  twice-iterated triangle inequality for the reals.
\item[(2)] One-sided threshold functions. Let real numbers
  $(t_{ab})_{a,b}$ satisfy $t_{\bar2\bar2}=\sum_{ab\not=\bar2\bar2}
  t_{ab}$ and define $f_{ab}(x)=[x\geq t_{ab}]$.  Note that this is an
  application of the shift in Thm.~\ref{thm:ftuple_closure} (D) with
  parameters $(-t_{ab})_{a,b}$ to the function-tuple $g_{ab}(x)=[x\geq
  0]$.
\item[(3)] Half-linear functions. For $t_{ab}$ as in (2), a
  positive slope $m$ and an exact constant function-tuple
  $c_{ab}$, define $f_{ab}(x) =
  \max(m(x-t_{ab}),c_{ab})$. That this tuple is in $\cT_4$ follows
  from the closure properties applied to the generating examples given
  above.
\item[(4)] Linear-edge window functions.  Choose thresholds
  $t_{l,ab}\leq t_{h,ab}$ such that for $d=l$ and for $d=h$,
  $t_{d,\bar2\bar2}=\sum_{ab\not=\bar2\bar2} t_{d,ab}$, and positive
  slopes $m_l$ and $m_h$. Define
  \begin{equation}
  \label{eq:linear_edge_window}
  f_{ab}(x)= \min(1,\max(0, m_h(x-t_{h,ab}),m_l(t_{l,ab}-x))).
  \end{equation}
  As illustrated in Fig.~\ref{fig:linear-edge_window}, these functions
  are $0$ between $t_{l,ab}$ and $t_{h,ab}$, and rise linearly away
  from these thresholds up to a value of $1$.  That they form tuples
  in $\cT_4$ follows from the closure properties.
\end{itemize}
In our applications, we use the linear-edge window functions.  In spot
checks using linear programming, they appear to optimize the sought
for violations among non-negative tuples in $\cT_4$ that are $1$
outside a fixed interval.

We now construct a $\CH$ function $l_{f}$ for pairs of timetag sequences
from an arbitrary function-tuple $f=(f_{ab})_{a,b\in \{1,2\}}$ in
$\cT_4$.  Let $\mathbf{r},\mathbf{t}$ be two ordered timetag
sequences, $\mathbf{r}=(r_1\leq ... \leq r_m)$ and
$\mathbf{t}=(t_1\leq ...\leq t_n)$.  Let $\cM$ be the family of
partial, non-crossing matchings between $\mathbf{r}$ and
$\mathbf{t}$. Such matchings $M$ can be identified with one-to-one,
partial, monotone functions $M:k\in\dom(M)\subseteq[m]\mapsto
M(k)\in[n]$.  Monotonicity implies that if $k,l\in\dom(M)$ and $k<l$,
then $M(k)<M(l)$.  (We use the notation $[j]=\{1,\ldots,j\}$.)  Let
$l_{f,ab}(\mathbf{r},\mathbf{t})$ be the minimum over all $M\in\cM$ of
the ``cost''
\begin{equation}
l(f_{ab},M,\mathbf{r},\mathbf{t})=m-|\dom(M)|+\sum_{k\in\dom(M)}f_{ab}(t_{M(k)}-r_k).
\label{eq:costdef}
\end{equation}
One way to think of this is as the minimum total cost of editing
$\mathbf{r}$ into $\mathbf{t}$ by deleting timetags in $\mathbf{r}$ at
a cost of $1$ (assessed by $m-|\dom(M)|$), deleting timetags of
$\mathbf{t}$ at no cost, and by shifting the remaining timetags of
$\mathbf{r}$ by $x$ at a cost of $f_{ab}(x)$ (assessed by the sum over
$k\in\dom(M)$), where each timetag can be shifted at most once and the
final time ordering is the same as the initial one.  We can also view
this as a maximum weighted bipartite non-crossing matching problem,
where the matching is between indices of $\mathbf{r}$ and indices of
$\mathbf{t}$ with the weight of $(k,l)$ being
$(1-f_{ab}(t_l-r_k))$. The cost is then given by $m$ minus the weight
of the maximum-weight matching.

\begin{Theorem}
  Suppose that the measurement outcome space $\cO$ consists of timetag
  sequences.  Let $f$ be a function-tuple in $\cT_4$. Then $l_f$ (as
  defined before Eq.~\eqref{eq:costdef}) is a $\CH$ function.
\label{thm:ftuple}
\end{Theorem}

\proof Consider deterministic LR outcomes $d^X_c$ as introduced
previously, but with outcomes consisting of timetag sequences.  Let
$M_{ab}$ be the cost-minimizing matchings for which
$l(f_{ab},M_{ab},d^A_a,d^B_b)=l_{f,ab}(d^A_a,d^B_b)$ if
$ab\not=\bar1\bar1$, and
$l(f_{\bar1\bar1},M_{\bar1\bar1},d^B_{\bar1},d^A_{\bar1})=l_{f,\bar1\bar1}(d^B_{\bar1},d^A_{\bar1})$.
We can construct a matching $M'$ from $d^A_{\bar2}$ to $d^B_{\bar2}$
by composing $M'=M_{\bar1\bar2}\circ M_{\bar1\bar1}\circ
M_{\bar2\bar1}$, with domain consisting of those indices for which the
composition is defined.  Then $M'$ is monotone and
$l(f_{\bar2\bar2},M',d^A_{\bar2},d^B_{\bar2})\geq
l_{f,\bar2\bar2}(d^A_{\bar2},d^B_{\bar2})$. Therefore, it suffices to
show that
\begin{equation}
l(f_{\bar2\bar2},M',d^A_{\bar2},d^B_{\bar2})\leq
 l(f_{\bar2\bar1},M_{\bar2\bar1},d^A_{\bar2},d^B_{\bar1})
 + l(f_{\bar1\bar1},M_{\bar1\bar1},d^B_{\bar1},d^A_{\bar1})
 + l(f_{\bar1\bar2},M_{\bar1\bar2},d^A_{\bar1},d^B_{\bar2}).
\label{eq:mmmm}
\end{equation}
The composition of functions defining $M'$ terminates at the first step
where the mapped-to index fails to be in the domain of the next
matching. This allows us to associate with each index not in the
domain of $M'$ a unique index along the way that is ``deleted'' in the
next step.  Indices in the domain pass through each matching and
accumulate separate distances that bound the corresponding distance in
$M'$.  To formalize this idea, for each index $k$ of $d^A_{\bar2}$, we
define $N(k)$ as a pair consisting of an index and a party/setting
label as follows: If $k\not\in\dom(M_{\bar2\bar1})$, then let
$N(k)=(k,A\bar2)$.  Else, if
$M_{\bar2\bar1}(k)\not\in\dom(M_{\bar1\bar1})$, then let
$N(k)=(M_{\bar2\bar1}(k),B\bar1)$. Else, if
$M_{\bar1\bar1}(M_{\bar2\bar1}(k))\not\in\dom(M_{\bar1\bar2})$, then
let $N(k)=(M_{\bar1\bar1}(M_{\bar2\bar1}(k)),A\bar1)$. If none of
these conditions apply, $k$ is in the domain of $M'$ and we let
$N(k)=(M'(k),B\bar2)$.  The definition implies that $N$ is one-to-one,
$k\in\dom(M')$ iff the second component of $N(k)$ is $B\bar2$, and if
$k\not\in\dom(M')$, then the first component of $N(k)$ is not in the
domain of one of $M_{\bar2\bar1}$, $M_{\bar1\bar1}$ or
$M_{\bar1\bar2}$. This ensures that all members of $d^A_{\bar2}$
deleted according to $M'$ are matched to deleted members of one of the
timetag sequences in the composition. In particular, the deletion cost
on the left-hand side of Eq.~\eqref{eq:mmmm} is at most that on the
right-hand side.

We now focus on the shift costs contributing to
Eq.~\eqref{eq:mmmm}. Let $(d^X_c)_k$ be the $k$'th timetag of $d^X_c$.
It remains to be shown that
\begin{eqnarray}
  \sum_{k\in\dom(M')}f_{\bar2\bar2}\left((d^B_{\bar2})_{M'(k)}-(d^A_{\bar2})_k\right)
  & \leq &
  \sum_{k\in\dom(M_{\bar2\bar1})}f_{\bar2\bar1}\left((d^B_{\bar1})_{M_{\bar2\bar1}(k)}-(d^A_{\bar2})_k\right) \nonumber\\
  &&\hspace*{.25in}
  +\sum_{k\in\dom(M_{\bar1\bar1})}f_{\bar1\bar1}\left((d^A_{\bar1})_{M_{\bar1\bar1}(k)}-(d^B_{\bar1})_k\right) \nonumber\\
  &&\hspace*{.25in}
  +\sum_{k\in\dom(M_{\bar1\bar2})}f_{\bar1\bar2}\left((d^B_{\bar1})_{M_{\bar2\bar1}(k)}-(d^A_{\bar2})_k\right).
\end{eqnarray}
Because all shift costs are positive, it suffices to show that
\begin{eqnarray}
  \sum_{k\in\dom(M')}f_{\bar2\bar2}\left((d^B_{\bar2})_{M'(k)}-(d^A_{\bar2})_k\right)
  & \leq &
  \sum_{k\in\dom(M')}\left[f_{\bar1\bar2}\big((d^B_{\bar2})_{M'(k)}-(d^A_{\bar1})_{M_{\bar1\bar1}\circ M_{\bar2\bar1}(k)}\big)\right.  \nonumber\\
  &&\hspace*{.25in}
  +f_{\bar1\bar1}\big((d^A_{\bar1})_{M_{\bar1\bar1}\circ M_{\bar2\bar1}(k)} -(d^B_{\bar1})_{M_{\bar2\bar1}(k)}\big) \nonumber\\
  &&\hspace*{.25in}
  +\left.f_{\bar2\bar1}\big((d^B_{\bar1})_{M_{\bar2\bar1}(k)} -(d^A_{\bar2})_k)\big)\right],
\end{eqnarray}
in which terms on the right-hand side that are not included in
the composed matchings have been neglected.  For each $k \in
\dom(M')$,
\begin{eqnarray}
  f_{\bar2\bar2}\left((d^B_{\bar2})_{M'(k)}-(d^A_{\bar2})_k\right) &=&
  f_{\bar2\bar2}\big((d^B_{\bar2})_{M'(k)}-(d^A_{\bar1})_{M_{\bar1\bar1}\circ M_{\bar2\bar1}(k)} \nonumber\\
  &&\hspace*{.25in} +(d^A_{\bar1})_{M_{\bar1\bar1}\circ M_{\bar2\bar1}(k)} -(d^B_{\bar1})_{M_{\bar2\bar1}(k)} \nonumber\\
  &&\hspace*{.25in} +(d^B_{\bar1})_{M_{\bar2\bar1}(k)} -(d^A_{\bar2})_k\big) \\
  &\leq &
  f_{\bar1\bar2}\big((d^B_{\bar2})_{M'(k)}-(d^A_{\bar1})_{M_{\bar1\bar1}\circ M_{\bar2\bar1}(k)}\big)  \nonumber\\
  &&\hspace*{.25in}+ f_{\bar1\bar1}\big((d^A_{\bar1})_{M_{\bar1\bar1}\circ M_{\bar2\bar1}(k)} -(d^B_{\bar1})_{M_{\bar2\bar1}(k)}\big)\nonumber\\
  &&\hspace*{.25in}+ f_{\bar2\bar1}\big((d^B_{\bar1})_{M_{\bar2\bar1}(k)} -(d^A_{\bar2})_k)\big),
\end{eqnarray}
according to the defining inequality of function-tuples.  These summands are
separate contributions (for distinct $k\in\dom(M')$) to the right-hand
side of Eq.~\eqref{eq:mmmm}, completing the proof.  \qed

An algorithm to compute the minimum costs
$l_{f,ab}(\mathbf{r},\mathbf{t})$ is described in
Appendix~\ref{sec:dyn}.

\section{Non-Signaling adjustments to $\CH$ functions}
\label{sec:nosigadj}

The non-signaling conditions are a set of equalities that constrain
probability distributions describing $A$'s and $B$'s measurement
outcomes.  They ensure that each party's outcome distribution is
independent of the other party's setting.  Otherwise, one party could
signal their setting to the other party.  Formally, given non-signaling
and any real-valued function $h$ on $\cO$, $\langle
h(O^A)|S^A=a,S^B=\bar1\rangle =\langle h(O^A)|S^A=a,S^B=\bar2\rangle$
and similarly for reversing the roles of $A$ and $B$.  Although both
quantum and LR theories obey the non-signaling equalities, using these
equalities to transform Bell inequalities can increase the
signal-to-noise ratio (SNR) of a Bell-inequality violation observed in
an experimental test of LR.

Consider the general iterated triangle inequality for a $\CH$ function
$l$.  We can modify $l$ by defining
\begin{equation}
l'_{ab}(x,y) = l_{ab}(x,y) + \left\{
  \begin{array}{ll}
     -f_{\bar1}(y)-g_{\bar1}(x)  & \mbox{if $a=\bar1$ and $b=\bar1$,}\\
      f_a(x)     +g_b(y) & \mbox{otherwise}.
  \end{array}\right.
\label{eq:nsadj}
\end{equation}
where $f_a$ and $g_b$ are arbitrary real-valued functions. Replacing
$l$ by $l'$ leaves the right-hand side of
Eq.~\eqref{eq:iterated_triangle2} unchanged, so $l'$ is also a $\CH$
function. Furthermore, we have $\langle B_l(T)\rangle = \langle
B_{l'}(T)\rangle$ for any model satisfying the non-signaling
conditions. We call $l'$ a non-signaling adjustment of $l$.  Note
that the functions
\begin{equation}
l_{A,ab}(x,y) = 
  \left\{\begin{array}{ll}
    -f_{\bar1}(y)& \textrm{if $a=\bar1$ and $b=\bar1$,}\\
     f_{a}(x) & \textrm{otherwise}
  \end{array}\right.
\end{equation}
and
\begin{equation}
l_{B,ab}(x,y) = 
  \left\{\begin{array}{ll}
    -g_{\bar1}(x)& \textrm{if $a=\bar1$ and $b=\bar1$,}\\
     g_{b}(y) & \textrm{otherwise}
  \end{array}\right.
\end{equation}
are $\CH$ functions, and $l'_{ab}=l_{ab}+l_{A,ab}+l_{B,ab}$.

Non-signaling adjustments can be used to improve the SNR of the
empirical estimate of a Bell function $B_l$ obtained from a sequence
of trials. As a simple example with a two-point outcome space,
consider $N$ trials to test a version of the inequality of
Eq.~\eqref{eq:bell_ineq}. The experiment is configured so that each trial
involves emission of exactly one photon pair with some probability
(and emission of nothing otherwise), and the measurement outcomes $1$
and $0$ correspond to whether a photon was detected or not. We start
with $l_{ab}(x,y)=\max(x-y,0)$, which is a $\CH$ function.  Let
$p^X_c$ be the probability that $X$ detects a photon at setting $c$ in
a trial.  Let $c_{ab}$ be the probability of coincident detections
when the parties use settings $ab$.  The expected value of the Bell
function of Eq.~\eqref{eq:chbelldef1} is then
\begin{eqnarray}
  \left.\begin{array}{l}
      \langle 4 [S^{A}=\bar2 \& S^{B}=\bar1] l_{\bar2\bar1}(O^{A},O^{B})\rangle\\
      {}+\langle 4 [S^{A}=\bar1 \& S^{B}=\bar1] l_{\bar1\bar1}(O^{B},O^{A})\rangle\\
      {}+\langle 4 [S^{A}=\bar1 \& S^{B}=\bar2] l_{\bar1\bar2}(O^{A},O^{B})\rangle\\
      {}-\langle 4 [S^{A}=\bar2 \& S^{B}=\bar2] l_{\bar2\bar2}(O^{A},O^{B})\rangle
    \end{array}\right\} &=&
  \left\{\begin{array}{l}
      \langle 4 [S^{A}=\bar2 \& S^{B}=\bar1] [O^{A}=1\&O^{B}=0]\rangle\\
      {}+\langle 4 [S^{A}=\bar1 \& S^{B}=\bar1] [O^{B}=1\&O^{A}=0]\rangle\\
      {}+\langle 4 [S^{A}=\bar1 \& S^{B}=\bar2] [O^{A}=1\&O^{B}=0]\rangle\\
      {}-\langle 4 [S^{A}=\bar2 \& S^{B}=\bar2] [O^{A}=1\&O^{B}=0]\rangle
    \end{array}\right.\nonumber\\
  &=&
  \left\{
    \begin{array}{l}
      p^A_{\bar2}-c_{\bar2\bar1}\\
      {}+p^B_{\bar1}-c_{\bar1\bar1}\\
      {}+p^A_{\bar1}-c_{\bar1\bar2}\\
      {}-(p^A_{\bar2}-c_{\bar2\bar2}).
    \end{array}
  \right. 
\end{eqnarray}
Note that the terms $p^A_{\bar 2}$ cancel in this expression.
However, when estimating the expectation by evaluating the Bell
function on the trials, the two $p^A_{\bar 2}$s are contributed by
values at different settings, namely $\bar2\bar1$ and
$\bar2\bar2$. Consequently, two sources of counting statistics
variability associated with detections of $A$ at setting $\bar2$
affect the the SNR of the Bell function.  We can eliminate this
problem by defining $f_{\bar2}(x)=-x$ in Eq.~\eqref{eq:nsadj}. (Here,
$f_{a}$ and $g_{b}$ are set to zero if not explicitly assigned.)  By
the non-signaling constraints and construction, the modified Bell
function has the same expectation as the original Bell function.  This
Bell function was used to show a Bell-inequality violation in the
experiment reported in Ref.~\cite{giustina:qc2013a}.  A further
improvement of the SNR is obtained by ``distributing'' the terms whose
expectations are $p^B_{\bar1}$ and $p^A_{\bar1}$ over the different
settings.  It is equivalent to averaging them over the other party's
setting choices and involves setting $f_{\bar1}(x) = -x/2$ and
$g_{\bar1}(x) = x/2$ in Eq.~\eqref{eq:nsadj}, in addition to setting
$f_{\bar2}(x)=-x$. This modification was introduced for the explicit
purpose of improving the SNR of the violation in
Ref.~\cite{christensen:qc2013a}.

\section{Protocol for data analysis}
\label{sec:analysis_protocol}

When analyzing a set of timetag-sequence pairs from trial measurement
outcomes, it is necessary to choose a Bell function $B_l$ that, before
the experiment, can be expected to show good results. It may not be
feasible to make a good choice of $B_l$ on purely theoretical
grounds. One can instead acquire a statistically useful training set
consisting of the outcomes from the first $N_t$ trials and set aside
the remainder in the analysis set. The notion of ``statistically
useful'' is not formalized here.  The training set is used to choose
the parameters required for analyzing the rest of the data.  The
training set is excluded from the final analysis. The main task is to
determine a function-tuple in $\cT_4$ and non-signaling adjustments to
use for defining $B_l$.  In principle, one can optimize the
function-tuple on the training set. That is, one can compute $B_l$ for
all such function-tuples and pick the one that minimizes the
empirically computed value of $B_l$. This optimization is difficult,
but one can use non-linear optimization on the $8$ independent
parameters of the linear-edge window functions. We note that on spot
checks, this subset of $\cT_4$ appears to contain the optimal solution
among function-tuples of $\cT_{4}$ whose members are constrained to be $1$
outside a fixed interval $[-u,u]$.  In the simulations below, the
number of independent parameters was reduced to $2$ by taking
advantage of symmetries.  The suggested non-linear optimization can
still be too resource intensive. In our simulations we used an
effective approximation; see Appendix~\ref{app:optimizing_tuples}. We
do not discuss methods for optimizing the non-signaling adjustments
here. In the simulations, we just use the ones described in the last
paragraph of Sect.~\ref{sec:nosigadj}. An even more ambitious
optimization could seek to optimize the SNR or the statistical
significance of the violation of the inequality rather than the
expected value of $B_l$.

\section{Bell function truncation for $p$-value bounds}
\label{sec:pbr}

Consider $N$ trials whose trial outcomes are $t_k$ and a Bell function
$B$. A direct way to analyze the trial outcomes is to compute $b_k =
B(t_k)$, let $f=\sum_k b_k$, and determine the sample standard error
for $f$ as $\sigma_e = \sqrt{N\sum_k(b_k-f/N)^2/(N-1)}$.  The
violation can then be quantified by the ``number of standard
deviations of violation'', $-f/\sigma_e$, which is the SNR of the
total violating signal.  This number needs to be interpreted with
care, particularly if it is very large--a desirable outcome of an
experiment. If the trials are i.i.d., then $f/N\pm\sigma_e/N$ is an
approximate 68 $\%$ confidence interval for the expectation of $B$.
(A better method that takes advantage of the fact that the settings
probability distribution is known is given in
Appendix~\ref{sect:nomsnr}.)  However, even in the case of
i.i.d. trials, $-f/\sigma_{e}$ does not quantify how strongly the
experiment ``rejects'' LR models. This is because the central limit
theorem cannot be reliably used to estimate extreme tail
probabilities. Furthermore, the assumption that the trials are
identical rarely holds to high precision, and independence cannot be
assumed in applications to cryptographic protocols.  For more details
on these issues, see~\cite{zhang_y:qc2011a,zhang_y:qc2013a}.

To determine the statistical significance of a Bell-inequality
violation, one can compute a bound on the largest probability with
which any LR model could produce a violation at least as large as that
observed.  This upper bounds a $p$-value according to the theory of
statistical hypothesis testing with respect to the composite null
hypothesis consisting of all possible LR models. Typical Bell
inequality experiments thus aim for extremely small $p$-values. Given
a $p$-value bound $p$, it is convenient to quantify the violation in
terms of the (negative) $\logp$-value bound, formally defined as
$-\log_2(p)$. In many cases, for example when reporting a discovery in
particle physics, $p$-values are converted to equivalent standard
deviations with respect to the one-sided tail probabilities of the
standard normal distribution.  For comparison, the $\logp$-values
corresponding to $1,2,3,4$ and $5$ standard deviations are
$2.7,5,9.5,14.9$ and $21.7$.

The first rigorous method for computing such a bound was given by
Gill~\cite{gill:qc2001a,gill:qc2003a} and is based on martingale
theory.  This ``martingale-based protocol'' does not require that
trials be independent from one trial to the next, or that they have
identically distributed measurement outcomes, an important feature for
its application to quantum randomness
expansion~\cite{pironio:qc2010a}.  The bound obtained by the
martingale-based protocol is suboptimal, but there is a protocol, the
PBR protocol~\cite{zhang_y:qc2011a}, that is optimal in an asymptotic
sense. Like the martingale-based protocol, the PBR protocol also does
not require independent or identical outcomes.  The PBR protocol has
the advantage that it does not require a predetermined Bell
inequality. Nor does it require that the number of trials be decided
in advance: It gives valid $p$-value bounds for any stopping rule (see
the last paragraph of this section). The full PBR protocol is
computationally infeasible when the measurement outcome spaces or the
number of settings are large, but there is a simplified and efficient
version of the protocol that still outperforms the martingale-based
protocol~\cite{zhang_y:qc2013a} while retaining the advantages of the
full protocol.

The PBR protocol is based on the following observation. Suppose that
before the $k$'th trial, we can determine a ``test factor'' $P_k$ on
the space of possible trial outcomes such that $P_{k}\geq 0$ and for
all LR models, $\langle P_k(T_k)\rangle_{\LR}\leq 1$, where the bound
holds regardless of what happened before the $k$'th trial in the
experiment. Then $P = \langle \prod_{k=1}^N P_k(T_k)\rangle_{\LR}\leq
1$.  Thus, given LR, according to the Markov inequality, the
probability that $P>1/p$ is bounded above by $p$, and therefore $1/P$
is a $p$-value bound. In general, candidate test factors $R$ can be
obtained from Bell functions $B$ bounded above by $z$ according to $R
= (z-B)/z$. More generally, given a collection of candidate test
factors $R_i\geq 0$ satisfying $\langle R_i(T)\rangle_{\LR}\leq 1$,
any convex combination of the $R_i$ can be used as $P_k$. (The
collection may depend on $k$.) The simplified PBR protocol takes as
input such a collection $(R_i)_{i=0}^{m}$ and chooses $P_k$ before the
$k$'th trial by optimizing the convex combination on the outcomes of
the previous trials. ($R_{0}$ is always chosen to be the ``trivial''
test factor $1$.) Specifically, it maximizes the empirical estimate of
the $\logp$-value increase per trial given by
$\sum_{i=1}^{k-1}-\log_2(P_k(t_i))/(k-1)$.  For more details, see
Ref.~\cite{zhang_y:qc2013a}.  The possibility of adjusting the test
factors for the upcoming trial after each trial makes it possible to
avoid making predetermined choices for the Bell function parameters.

In practice, there is little gained by re-optimizing the test factors
before every trial, and instead the factor is reused until
sufficiently many new trials have been obtained. In principle, the
number of trials before $P_k$ can be productively updated may be
determined from statistical considerations. (Our implementations so
far are largely based on heuristic considerations.) The simplest
method is to determine the optimal test factor from the training set
and use it uniformly on the analysis trials.  For example, the
starting Bell functions can include multiple choices of
parameters. The optimal convex combination of the corresponding test
factors can then be determined empirically on the training set and can
be directly applied to the trials in the analysis set.

To apply the PBR protocol to timetag data, we make two modifications
to the $\CH$ Bell functions. The first modification ensures that we
have a set of Bell functions that are bounded from above as required
by the simplified PBR protocol. Finite bounds are normally not
available for the timetag-sequence Bell functions discussed so far, or
the bounds are too high to be useful, so we describe a truncation
strategy below.  Our second modification increases the expected
$\logp$-value bound produced by the PBR protocol by shifting the $\CH$
functions so that the contributions to the violation of the $\CH$ Bell
inequality are equalized across measurement settings.

Consider a $\CH$ function $l$.  To obtain a bounded $\CH$ Bell
function, it suffices to modify $l$ by composition with a
function-tuple $f_{ab}$ in $\cT_4$ for which $f_{\bar2\bar2}$ is
monotone non-decreasing and the $f_{ab}$ are bounded. We call such an
$f_{ab}$ a monotone and bounded function-tuple.  To see that this
preserves the desired inequalities, consider
Eq.~\eqref{eq:iterated_triangle2} and define
$l'_{ab}(d^X_a,d^Y_b)=f_{ab}(l_{ab}(d^X_a,d^Y_b))$. Because of monotonicity of
$f_{\bar2\bar2}$, Thm.~\ref{thm:ftuple_closure} (F) ensures 
that $l'$ satisfies the inequality in
Eq.~\eqref{eq:iterated_triangle2}, thus $l'$ is also a $\CH$ function.

A convenient family of monotone and bounded function-tuples in $\cT_{4}$
is provided by
\begin{equation}
g_{ab}(x;b,u,c) = \min(\max(x+b_{ab},0),c)-u_{ab},
\label{eq:gabdef}
\end{equation}
where $b_{ab}$ and $u_{ab}$ are exact constant tuples in $\cT_4$ and
$c\geq 0$.
That $g$ is a function-tuple follows from the closure properties of
$\cT_{4}$ (Thm.~\ref{thm:ftuple_closure}).  If $l$ is modified to $l'$
by means of a function-tuple of the form $g_{ab}$, the Bell function
$B_{l'}$ is guaranteed to be bounded above by the maximum of
$c-u_{\bar1\bar1},c-u_{\bar1\bar2},c-u_{\bar2\bar1},u_{\bar2\bar2}$.
The following is a method for systematically choosing the truncation
parameters.

Consider a collection of training trials. Let
$\mathbf{l}_{ab}=(l_{ab,k})_{k=1}^{N_{ab}}$ consist of the observed
values of $l_{s^As^B}(o^A,o^B)$ for the trials where $s^A=a$ and
$s^B=b$. Let $\bar l_{ab} = \sum_{k} l_{ab,k}/N_{ab}$.  The bounds
$b_{ab}$ can be chosen as a compromise between having small bounds on
the Bell function and preserving the variation in the values of
$l_{ab,k}$. The violating signal is reduced if we truncate the values
of $l_{\bar2\bar2}$ so that the maximum value is too close to the mean
$\bar l_{\bar2\bar2}$.  This truncation point is determined by solving
$x_{\bar2\bar2}+b_{\bar2\bar2} = c$, that is
$x_{\bar2\bar2}=c-b_{\bar2\bar2}$.  (The upper or lower truncation
point is the value of $x$ for which $g_{ab}$ reaches its upper or
lower bound.)  Similarly, we should not truncate the other $l_{ab}$ so
that their minimum values are too close to their means $\bar l_{ab}$.
These truncation points are at $x_{ab}=-b_{ab}$.  Let $w_{ab}$ be a
``safe'' separation between the truncation points and $\bar l_{ab}$,
to be determined from the distributions of the $l_{ab,k}$ (see the end
of the next paragraph).  For $ab\not=\bar2\bar2$, we can set $b_{ab}$
by solving $-b_{ab}=\bar l_{ab}-w_{ab}$.  This determines
$b_{\bar2\bar2}=\sum_{ab\not=\bar2\bar2}b_{ab}$.  We can then choose
$c$ so that $c-b_{\bar2\bar2}=\bar l_{\bar2\bar2}+w_{\bar2\bar2}$.

The next step is to choose $u_{ab}$ so as to ensure that the
contributions to the violation conditional on the settings are
equalized.  This is done to improve the expectation of the
$\logp$-value bound by exploiting the concavity of the logarithm.  Let
$l'_{ab,k} = g_{ab}(l_{ab,k};b,0,c)$ with $b$ and $c$ as obtained so
far. Define $\bar l'_{ab}$ accordingly. We choose $u_{ab}$ so as to
balance the average violation for the different settings.  This is
accomplished by defining
\begin{equation}
u_{ab} = 
    \bar l'_{ab}-(-1)^{[a=\bar2\&b=\bar2]}(\bar l'_{\bar1\bar1}+\bar l'_{\bar1\bar2}+\bar l'_{\bar2\bar1}-\bar l'_{\bar2\bar2})/4,
\end{equation}
If we then use $g_{ab}$ as defined in Eq.~\eqref{eq:gabdef} and modify
$l$ as described there with $f_{ab}(x)=g_{ab}(x;b_{ab},u_{ab},c)$,
this ensures that each trial contributes the same estimated violation
$-v = (\bar l'_{\bar1\bar1}+\bar l'_{\bar1\bar2}+\bar
l'_{\bar2\bar1}-\bar l'_{\bar2\bar2})/4$ on average. If this violation
is not negative, then this truncation is not helpful for use in the
PBR protocol.  Assuming there is an empirical violation according to
the original $\bar l_{ab}$, the separations $w_{ab}$ need to be
increased until a violation remains visible in the $\bar l'_{ab}$.  In
this case, if the violation persists in future trials, the PBR
protocol can take advantage of the truncation.  In our implementation,
rather than attempting to find the optimal choice for $w_{ab}$, we
consider a small set of good candidates, obtain the associated Bell
functions and convert them to the non-trivial test factors $R_{i},
i\geq 1$ that are then convexly combined with the trivial test factor
by the simplified PBR protocol as described above. We found that the
convex combination chosen by the protocol normally involved more than
one choice of $w_{ab}$, suggesting that a single choice is not
optimal.

We remark that the shift $u_{ab}$ can be expressed as trivial
non-signaling adjustments by constant functions.  In addition to
increasing the SNR overall, a goal of non-signaling adjustments might
be to equalize the SNR conditional on the settings, but we did not
attempt to achieve this.

The truncation and shifting strategy above results in factors $P_{k}$
(and candidate test factors $R_i$) whose predicted expectations are
upper bounded by $4/3$, according to the training set.  To see this we
show that the upper bound $z$ of the modified Bell function that
determines a candidate test factor is at least $3v$, while by
construction, the predicted expectation is $-v$. Here, $v$ depends on
the training data and choice of $w_{ab}$ for this test factor.  Since
the test factor is given by $R=(z-B)/z$, we find that the predicted
expectation of $R$ is at most $(z+v)/z\leq 4/3$.  The upper bound is
given by
$z=\max_{ab,x}(-1)^{[a=\bar2\&b=\bar2]}g_{ab}(x;b_{ab},u_{ab},c)$.  The
expression for $g_{ab}$ shows that $\max_{x}g_{ab}(x;b_{ab},u_{ab},c)
= c-u_{ab}$ and
$\max_{x}-g_{\bar2\bar2}(x;b_{\bar2\bar2},u_{\bar2\bar2},c) =
u_{\bar2\bar2}$. Consider $u_{\bar2\bar2}$, which is the sum of the
other $u_{ab}$ (by definition of exact constant tuples and by
construction). For $ab\not=\bar2\bar2$, the lower bound on
$g_{ab}(x;b_{ab},u_{ab},c)$ is $-u_{ab}$. But the sample mean at
setting $ab$ of the truncated Bell function is $-v$, which requires
that its lower bound satisfies $-u_{ab}\leq -v$. It follows that
$u_{\bar2\bar2}\geq 3v$, which completes the argument.

A consequence of this observation is that the expected increase in the
$\logp$-value per trial is bounded by $\log_{2}(4/3)$, and it is not
possible to take advantage of seemingly strong violating signals per
trial.  To some extent, this is unavoidable: We are making no
assumptions on the probability distribution of the timetag sequences,
and an extremely adversarial LR model could take advantage of this in
future trials given our choice of parameters for the PBR protocol.
For the purpose of making the most of the PBR protocol, it is
therefore advantageous to design the individual trials to have
statistically small violating signals.  In particular, it is favorable
to have the one-trial SNR be well below $1$.  A simple way to
accomplish this and get a better overall $\logp$-value bound is to
shorten the durations of the trials and increase the number of trials
proportionally.

We finish this section by explaining our comment that the PBR protocol
can be used with any stopping rule, such as one according to which one
collects trials until a desired $p$-value bound is observed. To see
this, virtually replace the experiment with the stopping rule by one
that performs a fixed, large number of trials, larger than the maximum
number of trials that could be performed by the original
experiment. When the original experiment's stopping rule says
``stop'', the new experiment sets all future test factors to $1$. This
is justified because, the experimenter's choice of the test factors
for trial $k$ is only constrained by $P_{k}\geq 0$ and $\langle
P_k(T_k)\rangle_{\LR}\leq 1$, which are satisfied by $P_{k}=1$.  The
two experiments have the same statistics for the $p$-value bounds
obtained and the virtual experiment's $p$-value bounds are valid
according to the theory of the PBR protocol.  We remark that the PBR
protocol can be viewed as an application of the theory of
test-supermartingales as reviewed in Ref.~\cite{shafer:qc2009a}.

\section{Demonstrations on simulated data}
\label{sec:sim}

To illustrate timetag-sequence analysis we simulated experiments
intended to test inequalities such as Eq.~\eqref{eq:chbelldef1} and
its non-signaling variations. The situation is as described in
Sect.~\ref{sec:tt_bell_tests} with a Poisson source of
polarization-entangled photon pairs and high overall efficiency.  We
assume that both arms of the experiment have identical efficiency
$\eta$. In principle, such tests can succeed if
$\eta>2/3$~\cite{eberhard:qc1993a}. We explored the effects of uniform
jitter (defined in the next paragraph) on the performance of such
experiments at an efficiency of $\eta=0.8$. We also found lower bounds
on the maximum uniform and exponential jitter (defined in the next
paragraph) for which our techniques can show LR violation for
$\eta\geq 0.74$.  For each efficiency being considered, we first
optimized the violation of the CHSH inequality in Eq.~\eqref{eq:CHSH1}
by varying the settings choices and the parameter $\theta$ in the
family of unbalanced Bell states defined by
$\cos(\theta)\ket{00}+\sin(\theta)\ket{11}$. Given the efficiency, an
optimal state and settings, we computed the probabilities of the
measurement outcomes conditional on the settings.

We considered two families of timetag jitter distributions for the
difference between the recorded timetag and the true arrival time of a
photon. The first is the uniform distribution on an interval of width
$j_u$. That is, given the true arrival time $t$, the recorded timetag
$t'$ is uniformly distributed in the interval $[t,t+j_u]$.  The second
family is an exponential distribution with density
$\gamma e^{-\gamma(t'-t)}$ for $t'\geq t$. The two families of
distributions were chosen for ease of calculation and to illustrate
the effect of no tail versus long tail behavior, with long tails
leading to greater loss of violating signal.

The procedures for simulating and analyzing an experiment were
automated. We generated simulated photon pairs at a normalized rate of
$1$ per (arbitrary) unit of time. Thus the numerical value of $\tau$
(the mean photon-pair inter-arrival time) is $1$ in these units.  From
here on, time quantities such as $j_{u}$ are given as numerical values
with respect to these units.  The procedure is based on a choice of
observation window $T$ for the timetag sequences, number of training
trials $N_t$ and number of analysis trials $N_a$. In principle these
can be chosen before the simulation is started to ensure sufficient
data for determining the needed analysis parameters from the training
set. We recall that for the PBR analysis to take full advantage of the
SNR, it is a good idea to choose $T$ so that the SNR for one trial is
below $1$.  On the other hand, if $T$ is too small, loss of
coincidences near the boundary due to jitter leads to an additional
reduction of the violating signal. To reflect the conditions of
experiments with always-on pump lasers, we start generating photon
pairs $2$ units of time before the beginning of the observation
window. The data presented below uses $T=1000$, $N_t=10000$,
$N_a=200000$.  With these parameters and given the jitter
distribution, the procedure for generating and analyzing data is as
follows:

\begin{itemize}
\item[1.] Generate the trials for the training and analysis sets.  For
  each trial, first produce the sequence of times at which
  photon-pairs arrive at the detectors according to a Poisson process
  of rate $1$ as described above. The jitter distribution is
  used to delay the recorded time of detection independently for the
  two parties.  The timetags inside the observation window are saved
  to the parties' timetag sequences.
\item[2.] Determine the analysis parameters from the training set.
  Three studies are performed, each of which requires optimized
  parameters. The first is a conventional coincidence analysis, for
  which the coincidence window width is determined by optimizing the
  resulting violation on the training set. The second computes the
  timetag-sequence distances based on the linear-edge window
  function-tuples.  The function-tuple parameters are optimized on the
  training set.  The tuples are restricted to be reflection symmetric
  around $0$, and the thresholds for the settings other than
  $\bar2\bar2$ are taken to be identical. Thus only two parameters
  need to be optimized. (The optimization algorithm is described in
  Appendix~\ref{app:optimizing_tuples}).  The third is the PBR
  analysis on the truncated Bell function. The truncation method and
  optimization of the anticipated $\logp$-value increase per trial
  described in Sect.~\ref{sec:pbr} are used.
\item[3.] Perform the three analyses on the analysis set using the
  parameters determined from the training set.
\end{itemize}

The conventional coincidence analysis used in this procedure is
equivalent to using identical window functions with infinite
edge-slopes to compute the distances between each pair of timetag
sequences according to the prescription for using
function-tuples. Because the function-tuple effectively used is not in
$\cT_4$, there is no guarantee that the targeted Bell inequality is
strictly satisfied by LR models. The coincidence loophole examples
demonstrate that such an inequality requires additional assumptions on
the nature of the detection events. We remark that the analysis
described in \cite{larsson:qc2013a} is related to using
window functions with infinite edge-slopes, but the width of the
window function for setting $ab = \bar2\bar2$ is three times the width
of that used for the other settings. This set of window functions is
a function-tuple in $\cT_4$ and produces an analysis free of the
coincidence loophole.

\subsection{Uniform jitter distributions at efficiency $=0.8$}

\begin{figure}
{\hspace*{-1.5in}\includegraphics[width=7in]{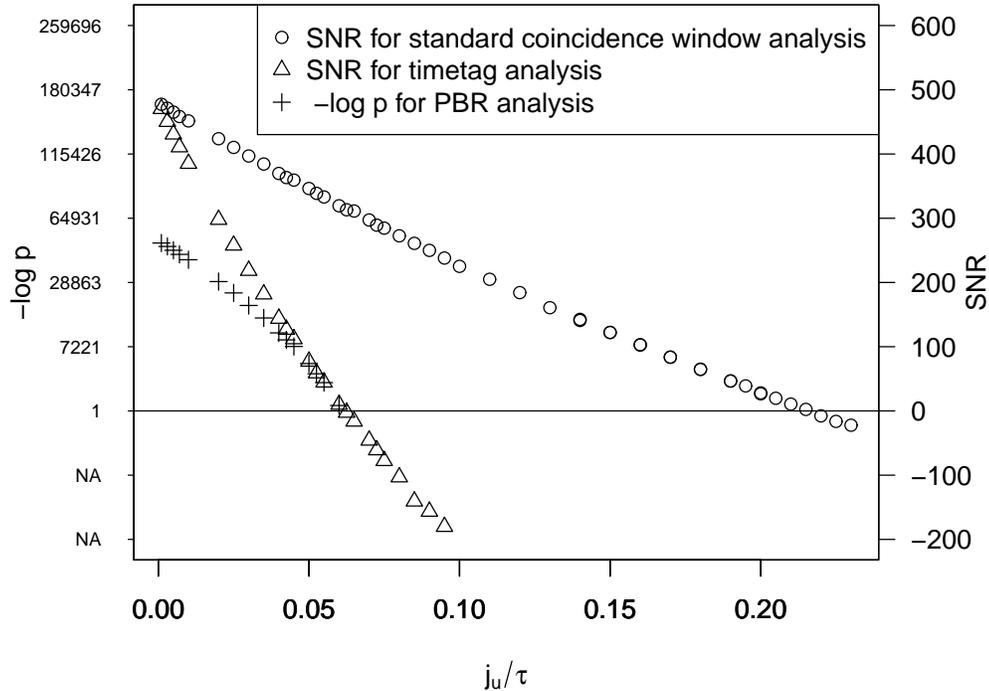}}
\caption{\label{fig:plot80a} Comparison of methods for simulated
  timetag data from a quantum source and detectors with efficiency
  $\eta=0.8$ and uniform jitter.  The nominal SNRs for the standard
  and the timetag analyses are shown on the right axis.  Negative SNRs
  mean that the signal is positive and therefore not violating.  The
  $\logp$-value bound for the PBR analysis is shown on the left axis.
  The horizontal axis shows the relative jitter width $j_{u}/\tau$.
  To match the two vertical axes with one another, we converted
  $\logp$-values to equivalent gaussian SNRs by computing the value
  for which the one-sided tail probability for the standard normal
  distribution matches the $p$-value bound. The computed $\logp$-value bounds
  are $0$ for $j_{u}$ above approximately $0.06$ and not shown on the plot.}
\end{figure}

We simulated experiments where the jitter has a uniform distribution
on the interval $[0,j_{u}]$, where $j_u\in[0.001,0.225]$.  This covers
most of the range for which the conventional coincidence analysis
shows a violating signal. Fig.~\ref{fig:plot80a} shows the nominal
SNRs for the conventional and the timetag analyses. The nominal SNR is
the ratio of the violating signal to the sample standard
deviation. Positive values are violating, negative ones are
non-violating.  We refer to this SNR as ``nominal'' because it cannot
be interpreted in terms of gaussian tail
distributions. See~\cite{zhang_y:qc2013a} for a discussion of this
issue. Our method for determining the SNRs is given in
Appendix~\ref{sect:nomsnr}.  The figure also shows the $\logp$-value
bounds from the PBR analysis on a matched scale. The $\logp$-value
bound and the timetag analysis' SNR both drop to zero around
$j_u=0.06$.

We next considered the question: for what $j_u$ does there exist an LR
source that has the same statistics as our simulated source?  We do
not have a definite answer to this question. However, we constructed
an LR source whose one- and two-point statistics closely match those
of the ideal Poisson source of quantum photon pairs for $j_u\geq
0.11$. More generally, we tweaked the LR source so that for all
positive values of $j_u$ it looks like the Poisson source that we
simulated for the data in Fig.~\ref{fig:plot80a}, except that it may
have more coincidences at the $\bar2\bar2$ setting depending on the
jitter. Further details are in
Appendix~\ref{app:lr_source}. Fig.~\ref{fig:plot80b} shows the results
from applying the conventional and timetag analysis methods to data
generated by a simulation of the LR source. As expected, the timetag
analysis shows no violation.  But the conventional analysis falsely
shows violation. For $j_{u}\geq 0.11$ the violation is similar if
somewhat lower than that for the corresponding quantum photon pairs in
Fig.~\ref{fig:plot80a}.

\begin{figure}
{\hspace*{-1.5in}\includegraphics[width=7in]{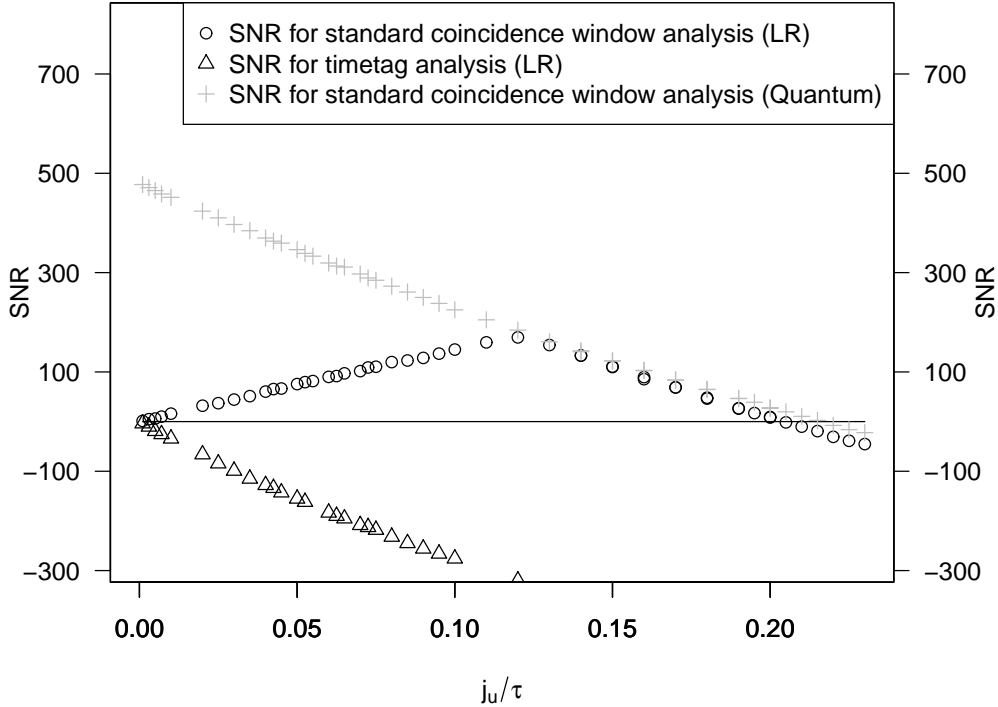}}
\caption{\label{fig:plot80b} Comparison of methods on LR-generated
  timetag data.  The conventional coincidence analysis shows a false
  violation over almost the entire range. The timetag analysis shows
  no violation. The $\logp$-value bounds from the PBR analysis are $0$
  everywhere and are not shown. The SNR of the Poisson quantum source
  whose one- and two-point statistics are approximated by the LR
  source for jitter larger than $j_u/\tau \geq 0.11$ is also shown.} 
\end{figure}

\subsection{Jitter thresholds for efficiency $\geq 0.74$}
\label{subsec:jw}

The simulations discussed above show that when there is too much
jitter, a nominally violating source of entangled photon pairs
produces timetag data that becomes indistinguishable from that
produced by an LR model.  While we cannot determine the minimum jitter
at which this happens, it is possible to lower bound the maximum
jitter at which the timetag analysis methods can see LR violation. We
simulated the photon-pair source and measurement configuration
introduced above at various efficiencies and varied the jitter
distribution.  We considered the uniform jitter model and the exponential
jitter model and determined the maximum jitter widths at which the
timetag analysis found violation.  The results are shown in
Tbl.~\ref{table:thresholds} in terms of the median of the jitter delay
of the recorded timetags.

\begin{table}
\centering
\begin{tabular}{c||c|c}
\hline
Efficiency & Uniform jitter & Exponential jitter\\
& median ($j_{u}/2$) & median\\
\hline\hline
0.74 & 0.013 & 0.0033
\\
0.76 & 0.018 & 0.0049
\\
0.78 & 0.024 & 0.0070
\\
0.80 & 0.031 & 0.0095
\\ 
0.85 & 0.052 & 0.017
\\
0.90 & 0.07  & 0.029
\\
0.95 & --- & 0.051
\\
\hline
\end{tabular}
\caption{Lower bounds on maximum jitter at which the timetag
  analysis can still detect violation of LR.  The rows give the
  maximum median jitter at which our simulations showed a
  violation as determined by the $\logp$-value being strictly
  positive. The
  simulation parameters other than jitter are the same as for the
  other simulations.  The jitter bounds are shown for the uniform and
  for the exponential distribution. For ease of comparison, they are
  parametrized in terms of the median delays of the recorded timetags.
  For low violation, unfeasibly large training and analysis sets are
  required to make the violation apparent in a simulation.  Thus, the
  entries in the table are lower bounds on the maximum jitter at which
  the timetag analysis can still detect violation.  The missing entry
  was not computed due to excessive computational resource
  requirements for our implementation. }
\label{table:thresholds}
\end{table}

\begin{acknowledgments}
  We thank Yi-Kai Liu and Krister Shalm for their help in revising
  this paper.  This work is a contribution of the National Institute
  of Standards and Technology and is not subject to U.S. copyright.
\end{acknowledgments}

\appendix

\section{Determining the violation's nominal SNR}
\label{sect:nomsnr}

The trials of an experiment result in a sequence of Bell function
values $(b_{i})_{i=1}^{N}$, one for each trial.  Let $s_{i}$ be the
settings of trial $i$ and $N_{ab}$ the number of trials at settings
$ab$. A standard approach to estimating the violation is to compute
the sample mean $\hat B_{ab} = \sum_{i:s_{i}=ab} b_{i}/N_{ab}$ and
define the estimated total violation as $N\sum_{ab} \hat
B_{ab}/4$. The variance of this value can then be estimated to first
order with respect to the variance of $N_{ab}$ from the sample
variances of the subsequences $(b_{i})_{i:s_{i}=ab}$. Instead of this
procedure, we used a method that makes no first-order
approximations and can be meaningfully applied even if the trials'
Bell function values $B_i$ are not independent.  We consider
this method better motivated, and it gives results that are
statistically close to those obtained by the standard approach.  Here
we describe the method for the case of i.i.d.  trials.

The goal is to estimate the expectation of the sum of the trials' Bell
function values and obtain a nearly tight bound on the variance of the
estimate.  The method is adaptive and applied to the analysis data
given some initial training data. (Calibration data or theoretical
predictions can be used instead.) The desired expectation is $\bar
B_{\textrm{tot}}=\sum_{i}\langle B_{i}\rangle$.  (Since we are
considering i.i.d. trials, the distribution of $B_{i}$ is the same for
all $i$.) Before considering the $i$'th trial, we use the previous
trials and the training set to obtain four unbiased estimates $\hat
B_{i,ab}$ of $\langle B_{i}|S_{i}=ab\rangle$, where $S_{i}$ is the
settings random variable for the $i$'th trial. This estimate can be
obtained by any means desired. We used the formula for $\hat B_{ab}$
given above and applied it to the training set and the first $i-1$
trials.  For the $i$'th trial, we then define the random variables
$F_{i} = \hat B_{i,S_i}$ and $\Delta_{i} = B_{i}-F_{i}$. Because the
settings probability distribution is known, so is the expectation of
$F_{i}$: $\langle F_{i}\rangle = \sum_{ab} \hat B_{i,ab}/4$.  We then
record $\delta_{i} = b_{i}-f_{i}$ for the $i$'th trial and continue.
Note that
\begin{eqnarray}
\sum_{i}\langle B_{i}\rangle &=& \sum_{i} \langle B_{i}-F_{i}\rangle
+\sum_{i}\langle F_{i}\rangle \nonumber\\
  &=& \sum_{i}\langle \Delta_{i}\rangle + \sum_{i,ab}\hat B_{i,ab}/4.
\end{eqnarray}
We can therefore empirically estimate $\bar B_{\textrm{tot}}$ as $\hat
B_{\textrm{tot}} = \sum_{i}\delta_{i} + \sum_{i,ab}\hat B_{i,ab}/4$,
which ensures that $\langle\hat B_{\textrm{tot}}\rangle = \bar
B_{\textrm{tot}}$.  We are considering the case of i.i.d. trials only,
so $\hat v = \sum_{i}\delta_{i}^{2}$ is an estimate of the variance of
$\hat B_{\textrm{tot}}$ that is biased high.  This is because the
variance of a random variable $W$ is the minimum of $\langle
(W-a)^{2}\rangle$ over $a$. Since $\Delta_{i}$ is designed to
asymptotically converge to a zero-mean random variable, the variance
estimate is asymptotically unbiased.  The SNRs for the conventional
coincidence analysis and the timetag-sequence analysis shown in
Figs.~\ref{fig:plot80a} and ~\ref{fig:plot80b} are defined as $\hat
B_{\textrm{tot}}/\sqrt{\hat v}$ multiplied by the sign of the
violation.

\section{Determining function-tuple-based distances}
\label{sec:dyn}

For general timetag-sequence pairs, computing a $\CH$ function $l_f$
(see Eq.~\eqref{eq:costdef}) can be accomplished by dynamic
programming.  The simplest implementation of this technique has a
quadratic time cost. For sequences such as those produced by our
simulations, which are associated with detections and coincidences
generated uniformly randomly in time, this cost can be reduced to
average linear time.

Let $d=(r_1\leq\ldots\leq r_m)$ and $e=(t_1\leq\ldots\leq t_n)$ be
timetag sequences. We wish to determine the distance $l_{f,ab}(d,e)$.
For this purpose, let $d_k=(r_1,\ldots,r_k)$ and
$e_l=(t_1,\ldots,t_l)$ be their initial segments. Let
$c(k,l)=l_{f,ab}(d_k,e_l)$. We can determine $c(k,l)$ inductively.  We
have $c(k,0)=k$ and $c(0,l)=0$. Suppose we have determined $c(k',l')$
for $k'<k$ and $l'\leq l$. Let $M_{k,l}$ be the (not-yet-known)
matching minimizing $l(f_{ab},M,d_k,e_l)$. There are two possibilities
to consider. Either $k\not\in\dom(M_{k,l})$, in which case
$c(k,l)=c(k-1,l)+1$; or $M_{k,l}(k)=l'\leq l$, in which case
$c(k,l)=c(k-1,l'-1)+f_{ab}(t_{l'}-r_{k})$. This reduction works
because $M_{k,l}$ is monotone.  Thus $c(k,l)$ can be determined as the
minimum of these possibilities.  In an algorithm, one can store the
$c(k,l)$ in an $(m+1)\times (n+1)$ matrix and fill its entries in the
order suggested by this inductive construction.  It is possible to
reduce memory requirements by filling the matrix row by row and
discarding the rows no longer needed. However, if it is desirable to
extract the cost-minimizing matching after $c(m,n)$ has been
determined, it helps to keep the full matrix, and work backward from
the $(m,n)$ entry to determine which of the cases above was used when
the entry encountered was computed.  The next entry as we work
backwards is determined by the case used, and the relationships in the
matchings arise from the encountered entries where the second case was
used.

The construction as given above can be too resource intensive for long
timetag sequences. For pairs of sequences with sufficiently low rates
of timetags, a simple way to speed up the algorithm is to break up the
sequences at sufficiently large gaps and apply the algorithm to each
resulting pair of subsequences. The subsequence costs are added.  To
formalize this idea, let $u>0$ be such that for all $x\not\in[-u,u]$
and for all $ab$, $f_{ab}(x)\geq 1$. In this case, no pair of indices
$k,l$ with $|t_l-r_k|\geq u$ needs to be considered for
matching. Break up the two sequences into pairs of subsequences
$d(k_{i,1},k_{i,2})=(r_{k_{i,1}}\leq\ldots\leq r_{k_{i,2}})$ and
$e(l_{i,1},l_{i,2})=(t_{l_{i,1}}\leq\ldots\leq t_{l_{i,2}})$ such that
$\max(d(k_{i,1},k_{i,2}),e(l_{i,1},l_{i,2}))+u\leq
\min(d(k_{i+1,1},k_{i+1,2}),e(l_{i+1,1},l_{i+1,2}))$ for every $i$ for
which both sides of the inequality are defined.  (The maximum of a
collection of sequences given as arguments to $\max$ is defined as the
maximum of the set of all timetags occurring in the arguments, and
similarly for $\min$.)  The algorithm is then applied to each
corresponding pair of subsequences and the costs computed are added to
determine the cost of the original sequence.

\section{Optimizing function-tuple parameters}
\label{app:optimizing_tuples}

When performing the timetag-sequence analysis on simulated data, we
optimized the parameters of the linear-edge window function-tuple
using a set of training data.  To simplify the optimization, we
restricted the linear-edge window functions in
Eq.~\eqref{eq:linear_edge_window} to be reflection symmetric around 0,
so that $m_l=m_h:=m$ and $t_{l,ab}=t_{h,ab}:=t_{ab}$.  We also fixed
$t_{\bar1\bar1}=t_{\bar1\bar2}=t_{\bar2\bar1}:=t$.  By definition of
the linear-edge window function-tuple, $t_{\bar2\bar2}=3t$, so only
the slope $m$ and truncation point $t$ remain to be optimized. Instead
of direct optimization, we used a simpler approximate optimization
strategy based on compressing the relevant information in the timetag
sequences.  This is accomplished as follows: For each pair of
training-set timetag sequences $\mathbf{r}$ and $\mathbf{t}$ at
setting $ab$ contributing to $B_l$, we first determine the optimum
matching $M$ for the ``compression'' function-tuple
$f_{ab}(x)=\min(\lambda |x|,1)$. In the simulations, $\lambda=1$.  In
general, $\lambda$ needs to be chosen so that the timetag differences
$\Delta$ for which the analysis function-tuple may be less than $1$
satisfy $f_{ab}(\Delta)<1$.  Our choice of $\lambda$ is given by the
photon-pair creation rate. But it should suffice to choose $\lambda$
so that few entangled photon pairs have recorded timetags that are
separated by more than $1/\lambda$.  For each index $k$ of
$\mathbf{r}$ in the domain of $M$, we determine the differences $x_k =
(t_{M(k)}-r_{k})$, where the notation for timetags of $\mathbf{r}$ and
$\mathbf{t}$ is as before. We then collect all such timetag
differences for all pairs $\mathbf{r}$ and $\mathbf{t}$ at setting
$ab$ in a single sequence $y_{ab}$. For the optimization, we assume
that the matchings $M$ obtained in the construction of $y_{ab}$ are
close in cost to the optimal matchings on the training set that would
be obtained according to function-tuples with the parameters we are
optimizing. Since we are working on the training, not the analysis
set, our computations can be approximate.  Given this assumption, we
can use $y_{ab}$ to compute an approximation of the $ab$ contribution
to the Bell function $B_{l_g}$ for a given function-tuple $g$.  Let
$X_{ab}$ be the maximum number of deletions that can contribute to the
$ab$-cost. This is given by the sum over all timetag-sequence pairs at
settings $ab$ in the training set of the number of timetags in the
first (when $ab\not=\bar1\bar1$) or second (when $ab=\bar1\bar1$)
timetag sequence in each pair of sequences. Then the estimated
difference between the $ab$-cost and $X_{ab}-|y_{ab}|$ is given by the
sum of the values of $g_{ab}$ applied to the timetag differences in
$y_{ab}$. (Here, $|y_{ab}|$ is the number of timetag differences in
$y_{ab}$.)  These sums can be computed much faster than one can
compute the exact minimum $ab$-cost for $g_{ab}$. The approximate
costs are used to optimize $g$, and the resulting $g$ is subsequently
used for the timetag analysis of the analysis set. We note that the
objective function in the optimization has irregularities that can
result in high sensitivity of the parameters of $g$ to statistical
noise.

\section{An LR source that exploits the coincidence loophole}
\label{app:lr_source}

In this section we describe the LR source whose false violation of a
Bell inequality under a standard coincidence window analysis is shown
in Fig.~\ref{fig:plot80b}.  This LR source can closely mimic the one-
and two-point statistics of a Poisson source of entangled photon pairs
detected by detectors with uniform jitter distribution of width
$j_u\geq 0.11$.

An LR source generates four timetag sequences $\mathbf{t}^X_{c}$,
where $X\in\{A,B\}$ and $c\in\{\bar1,\bar2\}$, for each trial.  These
sequences are the sequences that may be recorded by $A$ and $B$
depending on their settings. After the experiment, only the two
sequences corresponding to the actually chosen settings are visible to
the parties.  The goal is to match the visible statistics of the LR
source to those of a Poisson source of quantum photon pairs with
jitter.  We consider the uniform jitter distribution of width $j_u$
and a Poisson source whose detection statistics are determined by the
single-pair settings-conditional outcome probabilities $p(o^Ao^B|ab)$
and the uniform distribution for settings. We ensure that the LR
source exhibits the same marginal detection rates $p(o^{X}|ab)$.  But
since we are interested in how readily the conventional coincidence
analysis can be deceived, we allow the LR source to adjust the
apparent coincidence rates. Note that the inferred coincidence rates
depend on the method used to determine coincidences. The construction
of the LR source uses probabilities $p'(o^{A}o^{B}|ab)$ as a template
in an attempt to deceive the experimenter into believing $p'$. The
template satisfies $p'(o^{X}|ab)=p(o^{X}|ab)$. The apparent
coincidence rates are the same except at the $\bar2\bar2$ setting,
where we set $p'(11|\bar2\bar2) = p(11|\bar2\bar2)+\delta_c$,
$p'(00|\bar2\bar2) = p(00|\bar2\bar2)+\delta_c$, $p'(01|\bar2\bar2) =
p(01|\bar2\bar2)-\delta_c$, $p'(10|\bar2\bar2) =
p(10|\bar2\bar2)-\delta_c$, and $p'=p$ otherwise.  The coincidence
rate adjustment $\delta_c$ is chosen to maximize the rate at which the
LR source can successfully introduce an apparent violating signal.
For $j_{u}\gtrsim 0.11$, we found that it is possible to match the
coincidence rates ($\delta_{c}=0$).

\begin{figure}
\begin{center}
\hspace{0.03\textwidth}$a=\bar1$ \hspace{0.35\textwidth} $a=\bar2$\\
\includegraphics[width=0.4\textwidth, trim = 0cm 0cm 0cm 3.5cm, clip]{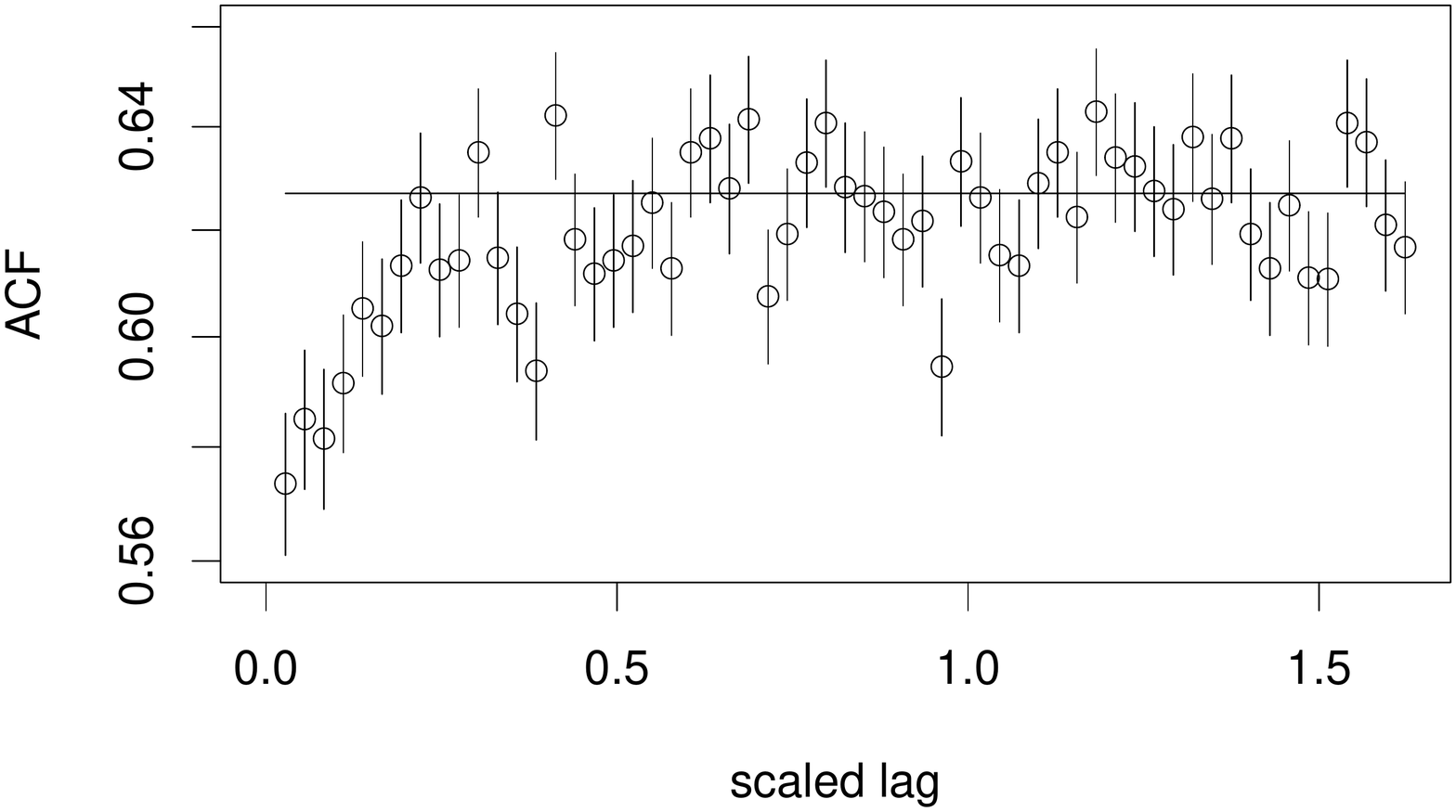}
\includegraphics[width=0.4\textwidth, trim = 0cm 0cm 0cm 3.5cm, clip]{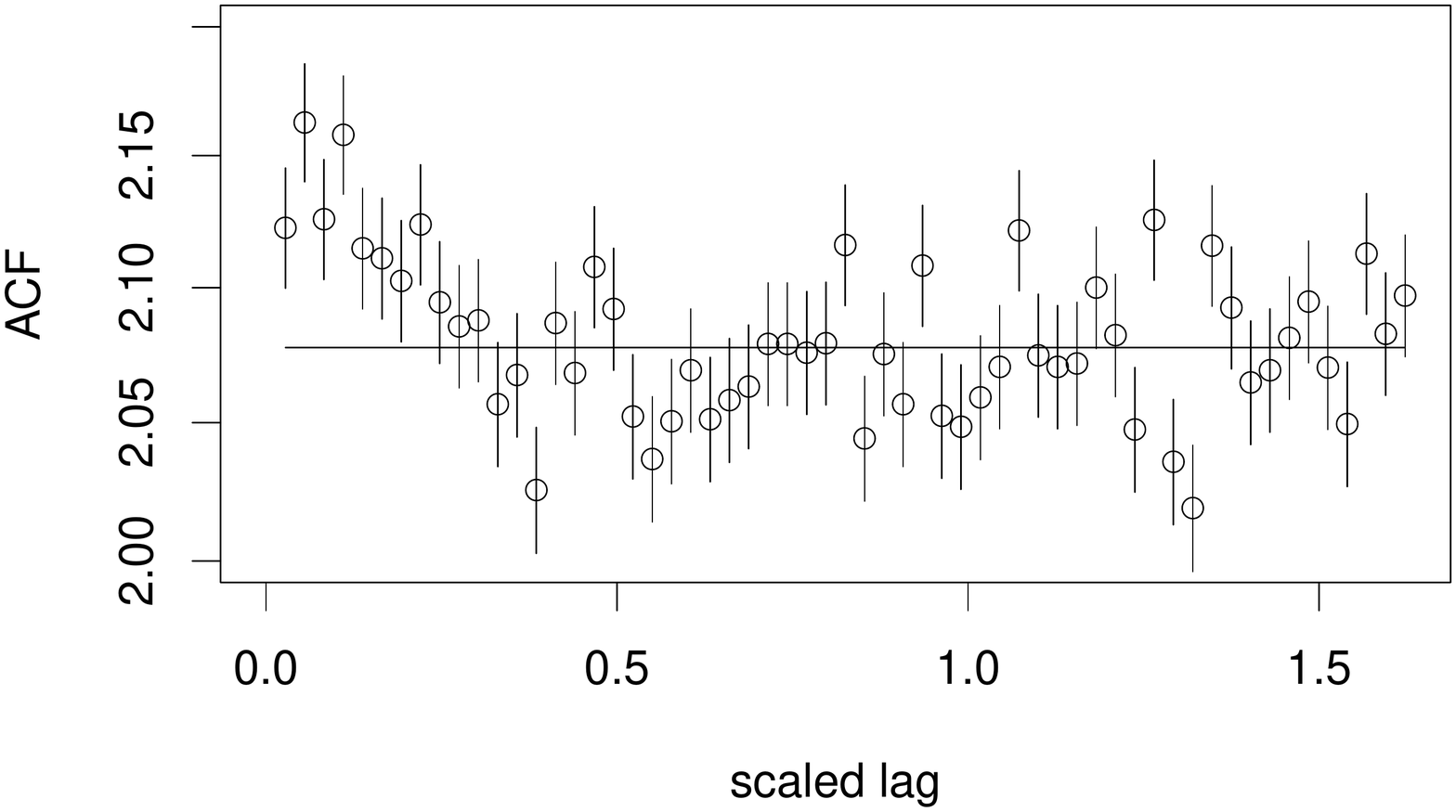}\\
\hspace{0.03\textwidth}$b=\bar1$ \hspace{0.35\textwidth} $b=\bar2$\\
\includegraphics[width=0.4\textwidth, trim = 0cm 0cm 0cm 3.5cm, clip]{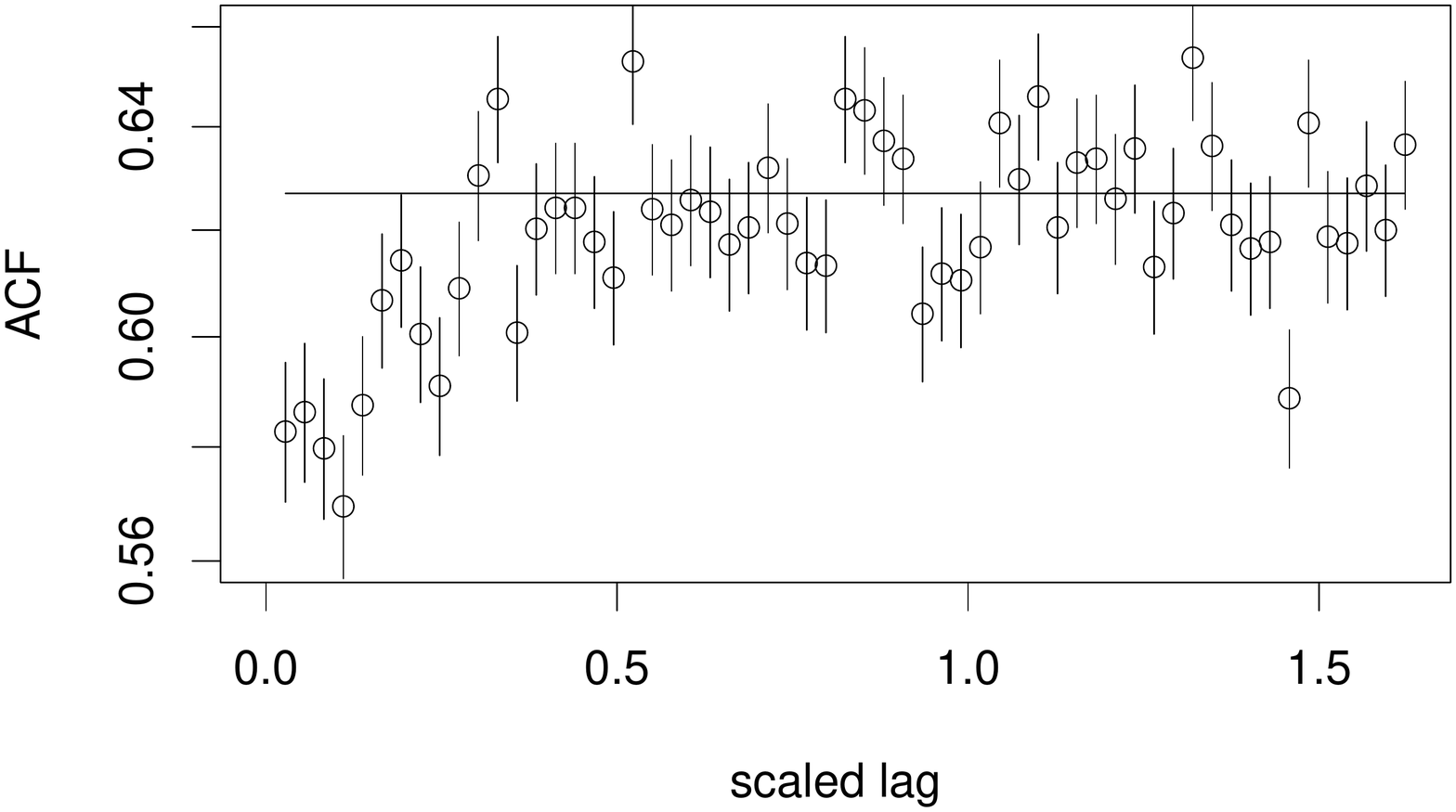}
\includegraphics[width=0.4\textwidth, trim = 0cm 0cm 0cm 3.5cm, clip]{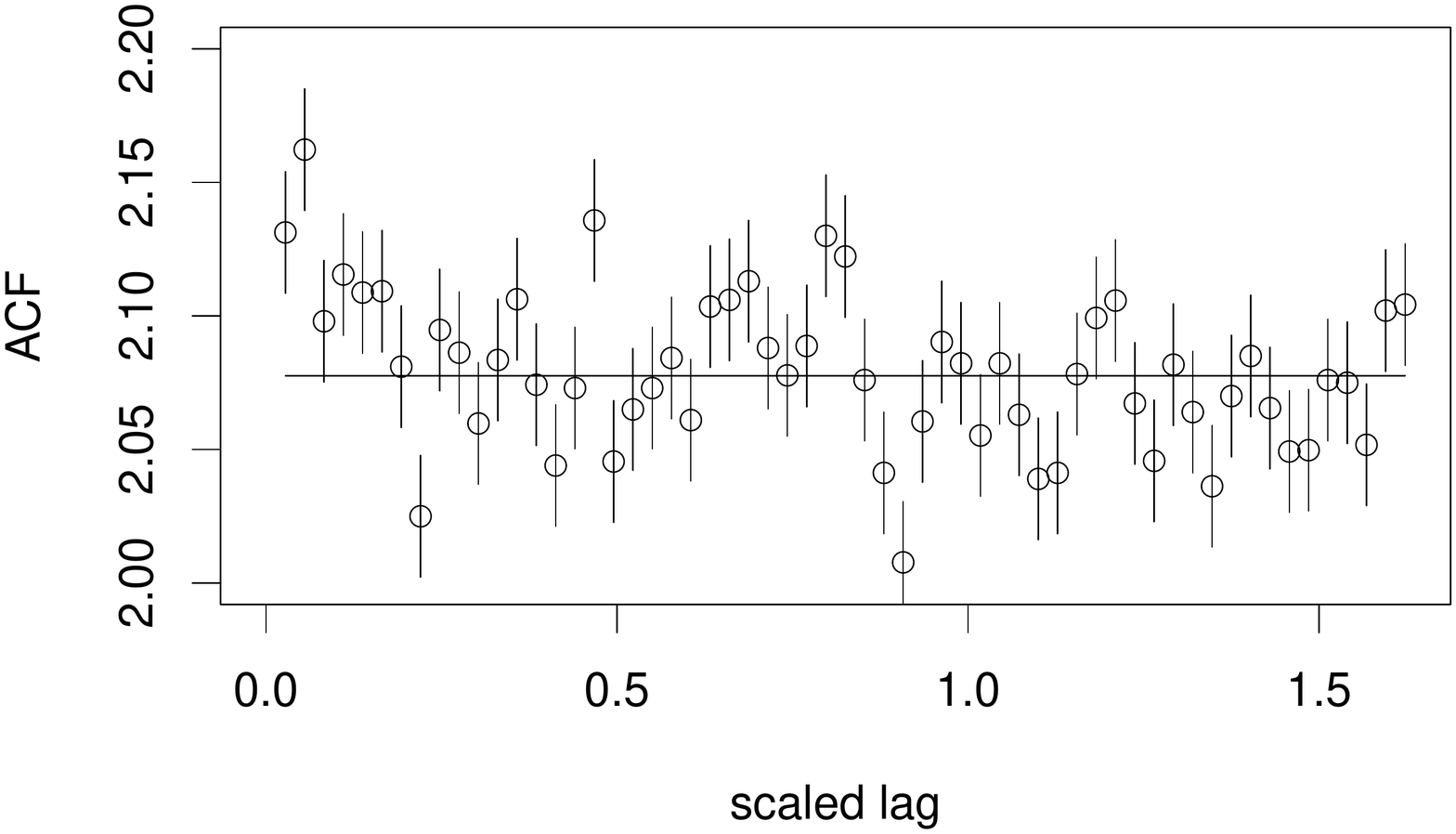}\\
\hspace{0.03\textwidth}$ab=\bar1\bar1$ \hspace{0.3\textwidth} $ab=\bar1\bar2$\\
\includegraphics[width=0.4\textwidth, trim = 0cm 0cm 0cm 3.5cm, clip]{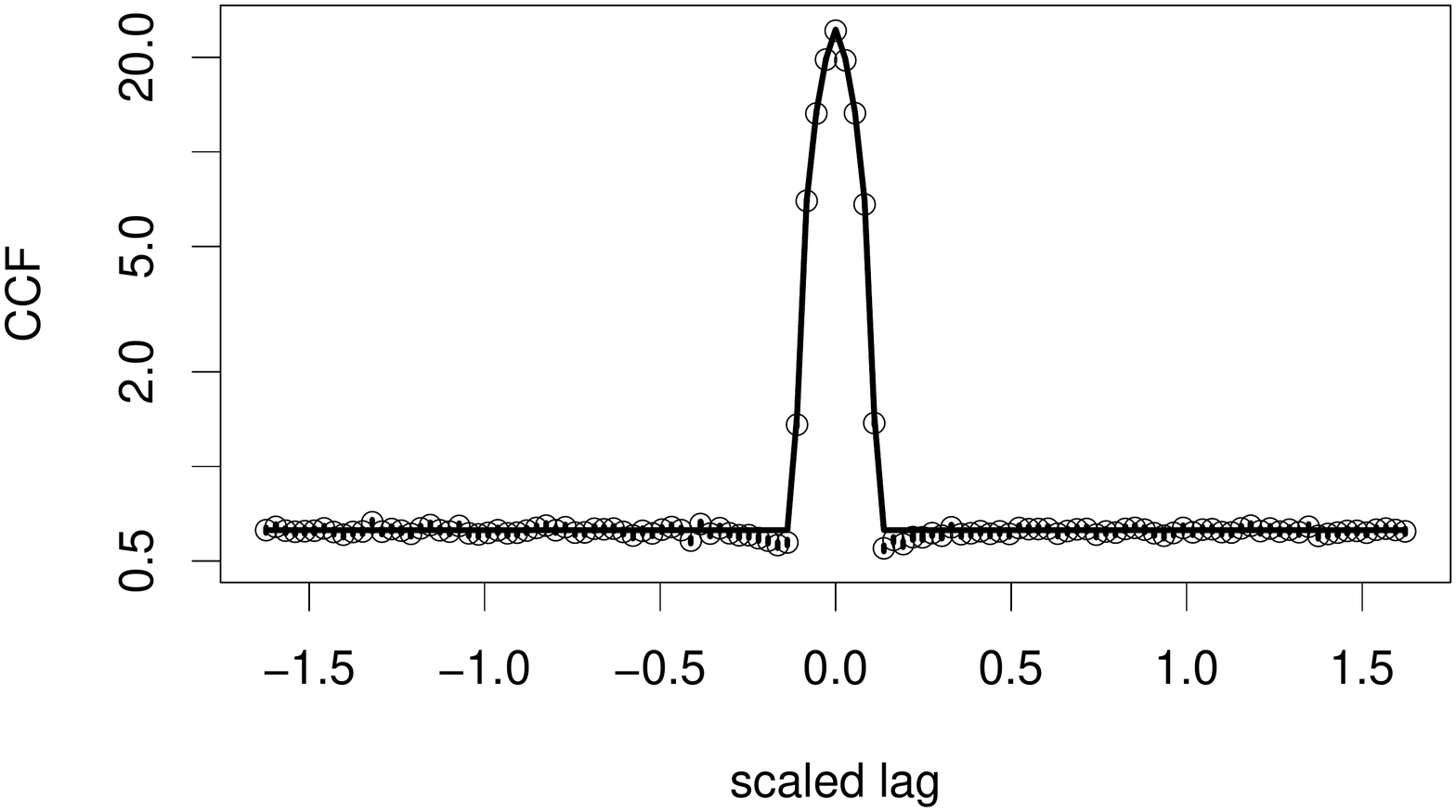}
\includegraphics[width=0.4\textwidth, trim = 0cm 0cm 0cm 3.5cm, clip]{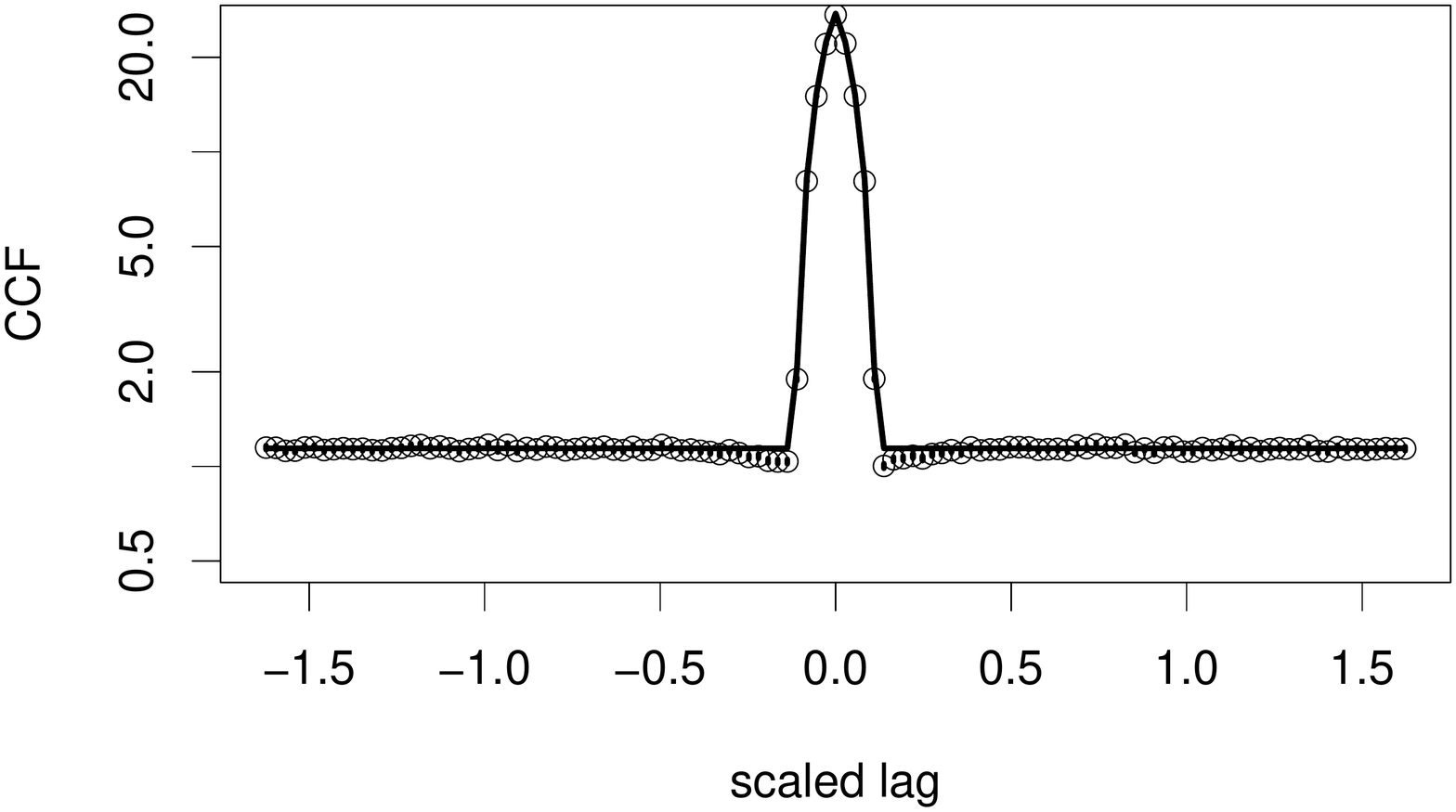}\\
\hspace{0.03\textwidth}$ab=\bar2\bar1$ \hspace{0.3\textwidth} $ab=\bar2\bar2$\\
\includegraphics[width=0.4\textwidth, trim = 0cm 0cm 0cm 3.5cm, clip]{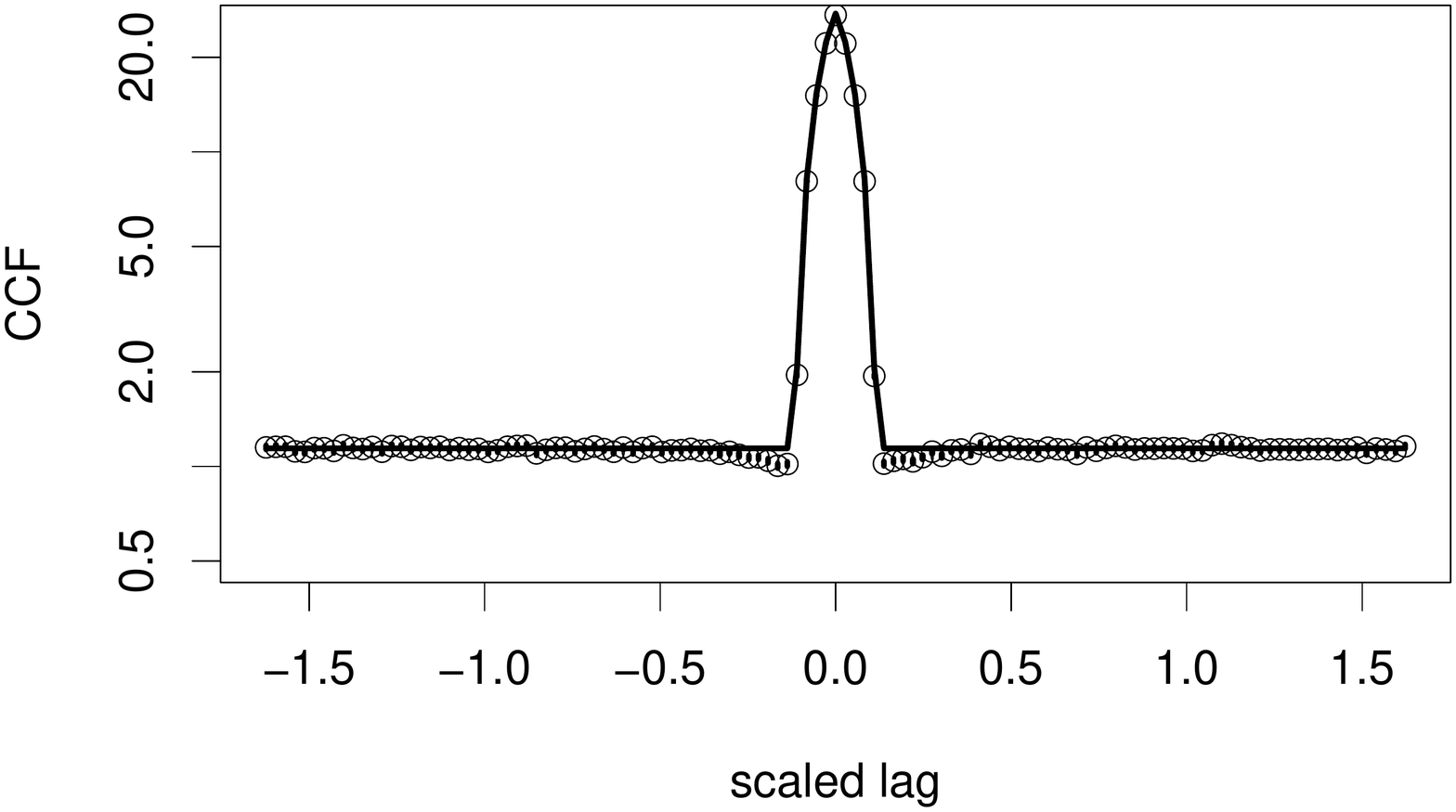}
\includegraphics[width=0.4\textwidth, trim = 0cm 0cm 0cm 3.5cm, clip]{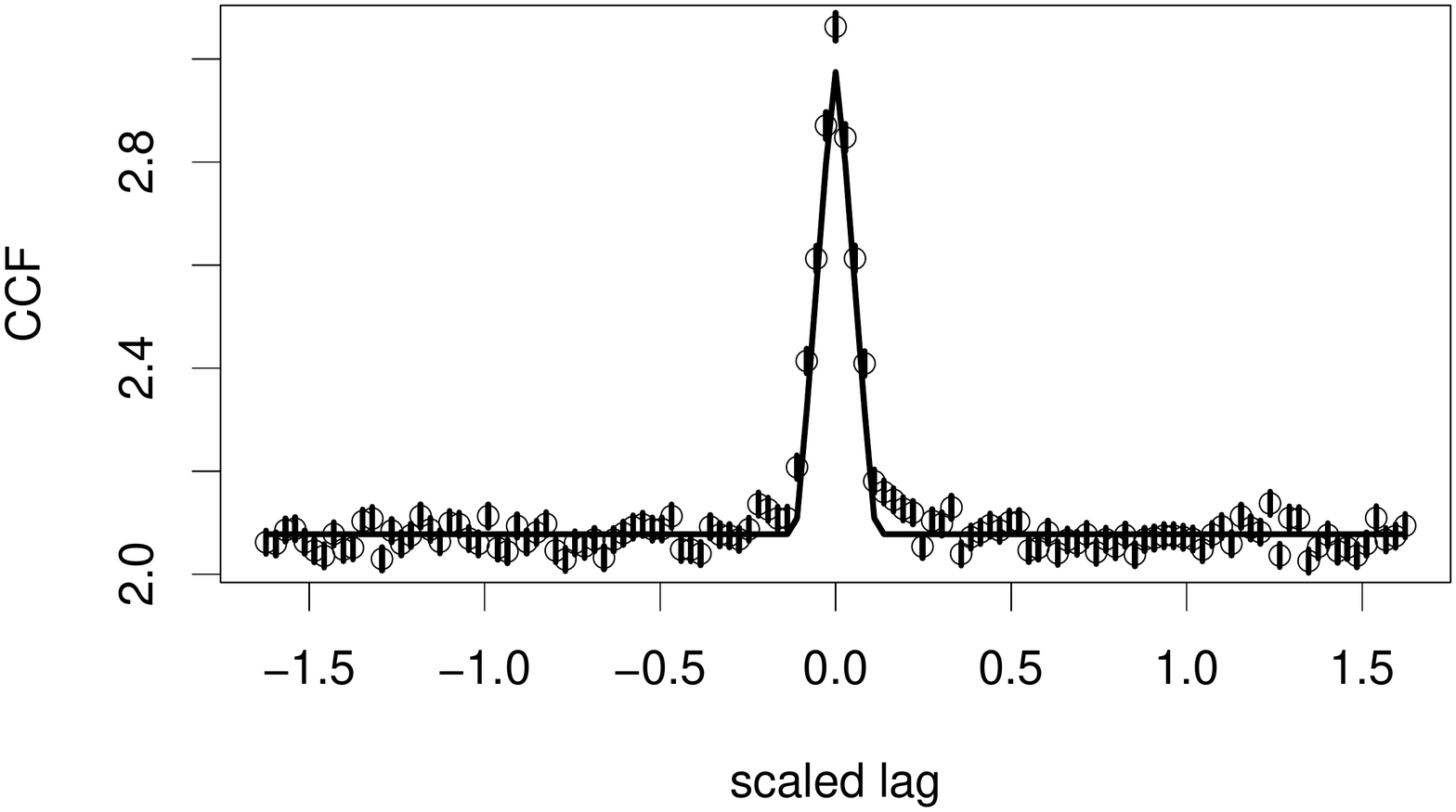}
\end{center}
\caption{\label{fig:plot80c} Auto- (ACF) and cross-correlation functions (CCF)
  (unnormalized) for LR-generated timetag data.  Each subfigure is
  labeled with the measurement setting(s) used to record the timetag
  data.  The points with error bars show estimated correlation
  function values and their associated approximate 68 \% confidence
  intervals for the LR timetag data determined from $200000$ trials.
  The LR source was designed to match timetag data from the quantum
  source with an efficiency $\eta=0.8$ and uniform jitter $j_{u}=0.11$
  that was used for the results shown in Fig.~\ref{fig:plot80a}.  The
  continuous curves are the corresponding theoretical correlation
  functions for the quantum source.  See the text for the definition
  of correlation functions used here. The bin width is $w_{b}=j_{u}/4$. The
  ``lag'' is defined as $d*w_{b}$.}
\end{figure}

For successful deception, the LR source's timetag-sequence statistics
should match that of a Poisson source with template frequencies given
by $p'$. (These frequencies account for the photon states, source
statistics and the jitter distribution.) We aimed for matching
detection rates and correlation functions. While our source does not
match the correlation functions exactly (see below), the residual
correlation mismatches are sufficiently small to either escape
detection or to be hidden by the typically much larger correlation
artifacts of the same order as the jitter introduced by the detection
apparatus. We also note that in an experiment, the source and jitter
distributions are not known beforehand, making a statistical test for
correlation artifacts introduced by the LR source difficult.
Comparisons of the correlation functions for $j_u=0.11$ are in
Fig.~\ref{fig:plot80c}.  Note that the mean time separation between
$\bar2\bar2$ coincidences seems to match that for other settings. That
is, there is no tell-tale broadening that might be expected from the
earlier coincidence-loophole examples in
Sect.~\ref{sec:tt_bell_tests}.

Here is a sketch of the LR source construction.  To generate LR
timetag sequences according to $p'$ and $j_u$, we first decompose $p'
= \lambda_{\lr} p_{\lr} + \lambda_{\pr} p_{\pr}$, where $\lambda_{\lr}$
is maximized subject to $p_{\lr}$ being an LR probability distribution
and $p_{\pr}$ a Popescu-Rohrlich (PR) box's~\cite{popescu:qc1997b}
probability distribution.  Here, the PR box is perfectly correlated on
all settings except at $\bar2\bar2$, where it is perfectly
anti-correlated.  To create an apparent PR box signal, we proceed as
follows.  Let $J(x,j_u)$ be the distribution of the time separation
$x$ between pairs of initially coincident timetags both of which are
jittered uniformly with width $j_u$. The density $J(x,j_{u})$ has a
triangle shape centered at $0$, supported on $[-j_{u},j_{u}]$, with
$J(0,j_{u})=1/j_{u}$. Let $p^X_{c}$ be the detection rate of $X$ at
setting $c$. We begin by generating timetags for $A$ at setting $\bar
2$ at a uniform rate $p^A_{\bar 2}$. This gives the timetag sequence
$\mathbf{t}^A_{\bar2}$. For each timetag $t$ thus generated, with
probability $\lambda_{\pr}/2$, we wish to generate a corresponding
timetag of $\mathbf{t}^B_{\bar2}$ that is generally far enough from
$t$ to be missed as a coincidence by the conventional analysis and is
indistinguishable from background.  The trick is to do this while
preserving the ability to generate $B$'s detections at setting $\bar
2$ according to a uniform process of rate $p^B_{\bar 2}$.  A first
attempt is to generate timetags for $\mathbf{t}^B_{\bar 2}$ so that
the rate of detections at $s$ is $\lambda_{\pr} J(s-t,3j_u)$ (three
times the apparent jitter width).  For fixed $t$, the probability of
at least one detection for $\mathbf{t}^B_{\bar2}$ thus generated is
$1-e^{-\lambda_{\pr}}$.  If there is at least one such detection, we
allocate it to a ``hidden'' coincidence for purposes of keeping track
of the detection statistics.  The hidden coincidence will be
attributed to a PR-like anticorrelation if this detection is not
recognized as a coincidence.

Note that detections generated at a uniform rate $r$ can be realized
by independently generating detections according to rate distributions
$\rho_i$ where $\sum_i\rho_i=r$.  Thus, if $\rho_1(s)=\sum_t
\lambda_{\pr} J(s-t,3j_u) \leq p^B_{\bar 2}$, we can generate further
detections to get the desired marginal statistics for $B$ at setting
$\bar 2$. Actually, we need to exclude the rate of coincidences,
$p'(11|\bar2\bar2)$ from $p^B_{\bar 2}$ in this inequality, because
these ``original'' coincidences are to be added separately later.  In
addition, to ensure that $\rho_1$ is below the corrected bound
$p^B_{\bar 2}-p'(11|\bar2\bar2)$, we have to deal with the problem
that for nearby timetags of $\textbf{t}^A_{\bar 2}$, the distributions
$J(s-t,3j_u)$ overlap. For this purpose we made
$\lambda_{\pr}=\lambda_{\pr}(t)$ depend on the timetag $t$ and used a
linear programming technique to maximize
$\sum_{t\in\mathbf{t}^A_{\bar2}}\lambda_{\pr}(t)$ subject to the
bound. The actual rate of apparent PR boxes is affected by the
result. These rates were determined by a Monte Carlo method as a
function of the jitter and matched to $\lambda_{\pr}/2$ by adjusting
$\delta_c$ as needed.

The PR boxes inserted into the timetags at setting $\bar2\bar2$ by
hiding coincidences need to be extended to recognizable coincidences
at the other settings. We first fill in these coincidences by
spreading them across the $\bar2\bar1$, $\bar1\bar1$ and $\bar1\bar2$
settings at the regular jitter width $j_{u}$ by dividing the hidden
coincidences' $\bar2\bar2$ separations equally into the separations
involving the timetags at the non-$\bar2$ settings.  Extra
coincidences with normal jitter width are then added to get the
desired coincidence rates at the non-$\bar2\bar2$ settings.  The
resulting marginal detection rates are subtracted from the uniform
rates $p^{X}_{\bar1}$ and extra detections are then filled in
accordingly.

While our implementation required substantial elaboration of the above
outline, the success of the strategy is witnessed by the false
violation discovered by the conventional coincidence analysis visible
in Fig.~\ref{fig:plot80b} and the empirical auto- and
cross-correlation functions shown in Fig.~\ref{fig:plot80c}.  The
figure shows the correlation functions for an LR source that simulates
the quantum sources used for the analyses shown in
Fig.~\ref{fig:plot80a} with $j_{u}=0.11$.  In this case, the LR
source's apparent coincidence rate at the $\bar2\bar2$ setting matches
the corresponding quantum source's rate. Small deviations from the
quantum source's correlation functions are visible in the plots. In
practice, such deviations are common and therefore hard to distinguish
from normal experimental artifacts.  Thus, we expect that it would be
difficult to find true LR violation in the quantum data, even if its
statistics cannot be exactly simulated by an LR source.  For
comparison, the maximum jitter for which the timetag analysis finds a
violation is about $j_{u}\approx 0.06$.

The estimated correlation functions shown in Fig.~\ref{fig:plot80c}
are determined from binned data as follows: Let $\mathbf{r}$ and
$\mathbf{t}$ be two timetag sequences and $w_{b}$ a bin width. Let $T$
be the total observation period of the timetag sequences and assume
that the initial time is $0$.  We construct functions
$b_{\mathbf{s}}:\{0,1,\ldots,N=\lceil T/w_{b}\rceil\}\rightarrow
\nats$ for $\mathbf{s}=\mathbf{r}$ and $\mathbf{s}=\mathbf{t}$ by
defining $b_{\mathbf{s}}(k)$ to be the number of timetags $s$ in
$\mathbf{s}$ with $kw_{b}\leq s<(k+1)w_{b}$.  The estimated
correlation function for $\mathbf{r}$ and $\mathbf{t}$ is defined by
$c(d,\mathbf{r},\mathbf{t}) =
\sum_{i=0}^{i+d=N}b_{\mathbf{r}}(i)b_{\mathbf{t}}(i+d)$.  (Note that
we do not normalize the correlation function.) The empirical values
shown in the figure are the sample means of the
$c(d,\mathbf{r},\mathbf{t})$ over the appropriate pairs of timetag
sequences from $200000$ trials. Note that the autocorrelation
functions are of the form $c(d,\mathbf{r},\mathbf{r})$ and are
symmetric about $d=0$, so only the half with $d\geq 0$ is shown.

We end with a brief note on how an adversary might realize the $\LR$
model of this section. Suppose that the adversary can surreptitiously
manipulate the measurement instruments of each party so that the
polarization angles of the two settings are approximately orthogonal.
Also suppose that the true jitter of the detectors is small but the
experimenter does not realize this. In this case, for each trial, the
adversary can randomly generate four timetag sequences according to
the $\LR$ model and, for each event in the timetag sequences, send a
photon at the event's time to the appropriate party with the
polarization that ensures it is only detected at the intended
setting. Because of the way that the $\LR$ model's timetag sequences
are generated, the detections will appear to be detections from
photon-pairs with detector jitter.  The manufacturer of the
measurement instruments can build-in the features that the adversary
needs to exploit in this scenario. It is therefore unlikely that a
protocol relying on a conventional coincidence analysis can achieve
device-independent security.  We leave open the question of whether an
LR model based on our techniques can be realized by an adversary who
can control only the source, when the characteristics of the
measurement instruments are fixed and known to the experimenter.


\end{document}